\documentclass[%
 reprint,
 amsmath,amssymb,
 aps,
]{revtex4-2}
\usepackage{aas_macros}

\usepackage{graphicx}
\usepackage{dcolumn}
\usepackage{bm}
\usepackage{xcolor}
\usepackage[normalem]{ulem}
\usepackage[colorlinks,urlcolor=blue,citecolor=blue,linkcolor=blue]{hyperref}
\usepackage[mathlines]{lineno}

\begin{document}

\preprint{APS/123-QED}

\title{Searching for structure in the binary black hole spin distribution}

\author{Jacob Golomb}
\affiliation{LIGO Laboratory, California Institute of Technology, Pasadena, CA 91125, USA}
\affiliation{Department of Physics, California Institute of Technology, Pasadena, CA 91125, USA}

\author{Colm Talbot}
\affiliation{LIGO Laboratory, Massachusetts Institute of Technology, 185 Albany St, Cambridge, MA 02139, USA}
\affiliation{Department of Physics and Kavli Institute for Astrophysics and Space Research, Massachusetts Institute of Technology, 77 Massachusetts Ave, Cambridge, MA 02139, USA}

\date{\today}

\begin{abstract}

The spins of black holes in merging binaries can reveal information related to the formation and evolution of these systems. Combining events to infer the astrophysical distribution of black hole spins allows us to determine the relative contribution from different formation scenarios to the population.
Many previous works have modelled spin population distributions using low-dimensional models with statistical or astrophysical motivations.
While these are valuable approaches when the observed population is small, they make strong assumptions about the shape of the underlying distribution and are highly susceptible to biases due to mismodeling. The results obtained with such parametric models are only valid if the allowed shape of the distribution is well-motivated (i.e. for astrophysical reasons). Unless the allowed shape of the distribution is well-motivated (i.e., for astrophysical reasons), results obtained with such models thus may exhibit systematic biases with respect to the true underlying astrophysical distribution, along with resulting uncertainties not being reflective of our true uncertainty in the astrophysical distribution
In this work, we relax these prior assumptions and model the spin distributions using a more data-driven approach, modelling these distributions with flexible cubic spline interpolants in order to allow for capturing structures that the parametric models cannot. We find that adding this flexibility to the model substantially increases the uncertainty in the inferred distributions, but find a general trend for lower support at high spin magnitude and a spin tilt distribution consistent with isotropic orientations. We infer that 62 - 87\% of black holes have spin magnitudes less than $a = 0.5$, and 27-50\% of black holes exhibit negative $\chi_{\rm eff}$. Using the inferred $\chi_{\rm eff}$ distribution, we place a conservative upper limit of 37\% for the contribution of hierarchical mergers to the astrophysical BBH population. Additionally, we find that artifacts from unconverged Monte Carlo integrals in the likelihood can manifest as spurious peaks and structures in inferred distributions, mandating the use of a sufficient number of samples when using Monte Carlo integration for population inference.
\end{abstract}

\maketitle

\section{Introduction}

Gravitational waves offer a unique probe into the properties of merging black holes (BHs) and neutron stars.
Since the first such detection in 2015, the LIGO-Virgo network \citep{LVKdetctors, LIGO, Virgo} has reported the detection of $\sim 90$ binary black hole (BBH) mergers, with each gravitational-wave (GW) signal encoding physical information about the BHs involved, such as their masses and angular momenta (spins) \citep{gwtc3, gwtc1, gwtc1}. Extracting this information has has enabled the study of properties of BBH systems on both an individual and population-level basis. From an astrophysical perspective, combining GW detections to infer the mass, spin, and redshift distributions of BBH systems can help answer questions ranging from binary formation and stellar evolution \citep{gwtc3pops, gwtc2pops} to the expansion rate of the Universe and possible deviations from General Relativity \citep{GWTC3Cosmology, GWTC3TGR}.

The spin of the BHs in a BBH system offer insight into the history of the binary. For example, BH spins can help reveal whether the BHs in a BBH system formed directly from core collapse of a heavy star or from the previous merger of two lighter BHs \citep{Kimball21, Fishbach22, GerosaFishbach, Fishbach17}. Although the processes governing the angular momentum transport out of a stellar core during collapse are not well-constrained, recent modeling work indicates that BHs resulting directly from core collapse supernovae should have negligible spin magnitudes \citep{FullerMa15, Qin, Heger}. While processes such as tidal interactions and mass transfer can induce higher spins on BHs in binary systems, it is uncertain how appreciable the resulting spin-ups can be \citep{Zevin22, Olejak, Bavera20, Bavera21}. On the other hand, BHs formed from the merger of two non-spinning BHs are expected to form a final BH with a relatively high spin magnitude \citep{Hinder, Hofmann, Fishbach17}, motivating the possibility to use spin magnitude as a tracer of a BHs formation history.

The direction of the BH spin vectors also encode information related to the formation history of a BBH system. Models suggest that BBH systems formed from common evolution, in which the component BHs evolve together from a stellar binary in an isolated environment free from significant dynamical interactions, should have component spin vectors nearly aligned with the orbital angular momentum axis, with any tilt being efficiently brought into alignment by tidal interactions \citep{Farr17, Kalogera2000}. On the other hand, BBH systems formed from dynamical encounters are not expected to have any correlated spins, such that the BH spin vectors are isotropic with respect to the orbital angular momentum \citep{Fishbach22, mandel10, Rodriguez}.

While only a couple of events in the third gravitational-wave transient catalog individually feature confidently high spin magnitudes or anti-alignment (i.e. a spin vector pointing opposite the angular momentum), hierarchically combining observations of GW events while folding in selection effects can reveal the degree to which these parts of spin parameter space contribute to the astrophysical distribution of BH spins. Previous work has used these inferred contributions to estimate the fraction of BBH systems in the local Universe which may have been formed hierarchically, dynamically, and by isolated evolution \citep{Doctor19, Fishbach22, Kimball21, gwtc3pops}. However, recent publications have disagreeing estimates for the contributions of anti-aligned and non-spinning BBHs to the astrophysical population. 

In \citep{gwtc3pops, gwtc2pops}, the authors conclude that the BBH distribution must feature anti-aligned spins at $> 90\%$ credibility, in contrast to the conclusion drawn in \cite{Roulet} that such anti-alignment is not evident in the population. In addition, \cite{Galaudage21} finds evidence for a non-spinning subpopulation of BHs, a conclusion which was challenged by \cite{Callister22}. While technical differences exist between works, a major possible contribution to some of these differing conclusions is model misspecification (see, e.g. \cite{Payne22, RomeroShaw22}); that is, different assumptions being imposed on the functional form of the spin distribution.

The \texttt{Default} model in \cite{gwtc3pops, gwtc2pops} models the distribution of the magnitude of the BH spin vector and the tilt angle between the spin vector and the orbital angular momentum. They adopt a Beta distribution for the spin magnitude model \citep{Wysocki19, gwtc3pops},
\begin{equation}\label{beta}
    \pi(a_{1,2}| \alpha_\chi, \beta_\chi) = \textrm{Beta}(a_{1,2}|\alpha_\chi, \beta_\chi),
\end{equation}
where $a_{1}$ ($a_{2}$) is the magnitude of the spin vector of the primary (secondary) BH, and $\alpha_\chi$ and $\beta_\chi$ are population hyperparameters determining the structure of the Beta distribution. The model for the distribution of tilt angles, $\theta$, is motivated by two subpopulations: one preferentially aligned ($\cos(\theta) \approx 1$) and one isotropic \citep{Talbotspin, gwtc2pops, gwtc3pops}. The model is parameterized as:
\begin{equation}\label{defaulttilts}
    \pi(\cos\theta_{1,2}|\xi, \sigma_t) = \xi G_{t}(\cos\theta_1|\sigma_t)G_{t}(\cos\theta_2|\sigma_t) + \frac{1 - \xi}{4},
\end{equation}
where $G_t$ is a truncated Gaussian centered at $\cos\theta = 1$ with standard deviation $\sigma_t$ and bounded in $[-1, 1]$, and $\xi$ is the relative mixing fraction between the subpopulations. The second term corresponds to the contribution from the uniform (isotropic) distribution. 

This population model has been extended in other work to allow for other astrophysically-motivated features to help draw conclusions related to the different formation scenarios present in the astrophysical distribution \citep{Galaudage21, Roulet, Callister22, Vitale2022}. Adopting an astrophysically-motivated, strongly parametric model necessarily limits the possible features resolvable in the inferred distribution to what the chosen function can model. Accordingly, in this work, we consider a strongly parametric model to be one that has a specific, prior-determined shape as provided by the parameterization (e.g. a normal distribution), which is then constrained by the data. When using such a distribution to draw astrophysical conclusions from the inferred population, this is a reasonable and intended consequence, as the model is chosen to encode prior beliefs on the parameters that should govern the astrophysical distribution; however, if additional features exist in the true astrophysical distribution and a strongly parametric model cannot account for them, such features can be missed and a biased result may be obtained.

Previous work has shown that substructures in the BH mass distribution can be captured by cubic splines acting as a perturbation on top of a simpler parameteric model \citep{Edelman, gwtc3pops}. In \cite{Edelman}, the authors consider an exponentiated spline perturbation modulating an underlying power law in the mass distribution. In this work, we model the spin magnitude and tilt distributions using exponentiated cubic splines modulating a flat distribution to obtain a more data-driven result for the inferred population of BH spins. In doing so, we limit the potential bias caused by mismodeling the spin distribution and allow for the possibility of capturing features not accessible with a strongly parametric model. 

The remainder of the paper is organized as follows. In Section \ref{models}, we detail the functional form and implementation of the cubic spline model. We provide the background of hierarchical Bayesian inference in Section \ref{methods}, as it applies to population inference with GW sources. In Section \ref{results} we present the resulting spin distributions we obtain for various spline models adopted in this work. Finally, we use these results to draw conclusions related to the astrophysical distribution of BBH spins and provide a relevant discussion in Section \ref{discussion}.
We additionally supply three appendices; the first provides additional details about an efficient caching technique for the cubic spline model, the second explores the effect of uncertainty in our estimation of the selection function, and the third describes robustness of our results to different choices of prior distribution.

\section{Models}\label{models}

Following the model for the black hole mass distribution considered in~\cite{Edelman}, we fit the distribution of spin magnitudes and cosine tilts using exponentiated cubic splines
\begin{equation}\label{spline_exp}
    p(x) \propto e^{f(x)}.
\end{equation}
A spline is a piecewise polynomial function defined by a set of node positions, the value of the function at those nodes, and boundary conditions at the end nodes.
We use a cubic spline as it is the lowest order spline that enforces continuity of the function and its first derivative everywhere.

\subsection{Node positions and amplitudes}\label{sec:splinepriors}

In this work we consider models with 4, 6, 8, and 10 nodes spaced linearly in the domain of the parameter space. For the two distributions being modelled with splines, this gives 16 unique spin models (4 amplitude node placement models $\times$ 4 tilt node placement models).
Our choice for the prior on the amplitude of each node is a unit Gaussian distribution.
Comparisons with other node amplitude prior choices are detailed in Appendix \ref{appendix:prior}.

In order to fully characterize a cubic spline, the first and second derivatives must be determined at each node.
For all but the endpoints, these derivatives are specified by requiring continuity in the spline and its derivative.
At the endpoints, there is no unique way to determine this and a range of boundary conditions are commonly used.
For our implementation, we want the prior distribution of the derivatives at the endpoints to match that of the internal nodes.
This requires providing two additional free parameters at each end of the spline.
In practice, we add two additional nodes outside each boundary, with amplitudes that are free to vary according to the prior. Throughout this work, the number of nodes in a model refers to the number of nodes within the domain (i.e. not including these outside nodes). The spacing between these nodes is the same as that between nodes within the domain. 

\subsection{Modeling spins with splines}

In this work, we use the spline model detailed above to model the population of spin magnitudes $a$ and tilt angles $\cos\theta$.
Consistent with \cite{gwtc2pops, gwtc3pops}, we model these parameters as independent and identically distributed.
The total spin population model is
\begin{equation}
    \pi_{\rm spin}(\eta|\Lambda_s) = p_{a}(a_1) p_{a}(a_2) p_{t}(\cos\theta_1) p_{t}(\cos\theta_2),
\end{equation}
where $\Lambda_{s}$ is the set of population hyperparameters controlling the spline node location and amplitudes.
The functions $p$ are determined from Eq.~\ref{spline_exp}.
The domain of the spin magnitude distribution extends over $a \in [0, 1]$ and that of the spin tilt distribution covers $\cos\theta \in [-1, 1]$.

\section{Methods}\label{methods}

\subsection{Hierarchical Bayesian Inference}
In order to constrain the spin magnitude and tilt distribution, we carry out hierarchical Bayesian inference in which we calculate the likelihood of the entire observed dataset given a set of population hyperparameters $\Lambda$ while marginalizing over the uncertainty in the physical parameters of each event.
After analytically marginalizing over the total merger rate $R$ with a prior $\pi(R) \propto R^{-1}$, we express the likelihood of the hyperparameters $\Lambda$ parameterizing the population is expressed as (e.g. \cite{thranetalbot19}):
\begin{equation}\label{likelihood}
    \mathcal{L}(\{d\}|\Lambda) \propto p_{\text{det}}(\Lambda)^{-N} \prod^{N}_i \int \mathcal{L}(d_i|\eta_{i}) \pi(\eta_{i}|\Lambda)d\eta_{i}. 
\end{equation}
Here, $\mathcal{L}(d_i|\eta_i)$ is the likelihood of observing the data $d$ from the $i$th event, given physical (i.e. single-event) parameters $\eta_i$. In this work, $\eta_i$ consists of masses, spins, and redshift of the $i$th event.
The quantity $p_{\text{det}}(\Lambda)$ encodes the sensitivity of the search algorithm that identified the signals and is described in more detail in Sec.~\ref{sec:selection}.

Our population model $\pi(\eta|\Lambda)$ describes the astrophysical distribution of masses, redshifts, and spins. We model the primary mass distribution with the Powerlaw + Peak model \citep{TalbotPowerLawPeak}, the mass ratio ($q$ $ =\frac{ m_2}{m_1}$) distribution with a power law, and the redshift distribution also with a power law, with source-frame comoving merger rate density $R(z) \propto (1 + z)^3$ \citep{Fishbach18, gwtc3pops, gwtc2pops}. We choose to fix the redshift distribution because we use our own injection set to estimate sensitivity, thresholding on SNR rather than FAR to determine ``found" injections. Since this makes the threshold used to select real events (FAR $<$ 1 yr) slightly different from that used to threshold sensitivity injections, and the redshift distribution is particularly sensitive to the near-threshold events, we fix the redshift distribution in order to avoid biases (see \cite{gwtc1pops} for an example of where a similar approximation was used). See Section \ref{sec:selection} for details on sensitivity injections.  We list the hyperparameters $\Lambda$ and their corresponding priors in Table \ref{tab:priors}.

\begin{table}[]
    \centering
    \begin{tabular}{||c c c||} 
     \hline
     Parameter & Description & Prior \\ [0.5ex] 
     \hline\hline
     $\alpha$ & Power Law index for $m_1$ & (-4, 12) \\ 
     \hline
     $\beta$ & Power Law index for $q$ & (-2, 7) \\
     \hline
     $M_{\max}$ & Maximum mass & (60, 100) \\
     \hline
     $M_{\min}$ & Minimum mass & (2, 7) \\
     \hline
     $\lambda$ & Fraction of sources in Gaussian peak & (0, 1) \\ 
     \hline
     $M_{pp}$ & Location of Gaussian peak & (20, 50) \\
     \hline
     $\sigma_{pp}$ & Standard deviation of Gaussian peak & (1, 10) \\
     \hline
     $\delta_{m}$ & Minimum mass turn-on length & (0, 10) \\ [1ex] 
     \hline
    \end{tabular}
    \caption{Priors for mass distribution used in hierarchical inference, consistent with those used in \cite{gwtc3pops}. Priors on the spin distribution are described in Section \ref{models}. Priors are uniform between the bounds listed in the third column. }
    \label{tab:priors}
\end{table}

An initial choice that must be made when computing Eq.~\ref{likelihood} is which events to include in the analysis. Typically this is done by establishing some detection threshold on the Signal-to-Noise Ratio (SNR) or False Alarm Rate (FAR) and including all events that pass this threshold. We choose to include the 59 events in the third observing run (O3) of the LIGO-Virgo network which have a False Alarm Rate of less than 1 year$^{-1}$ and are included in the main BBH analysis of \citep{gwtc3pops}. We limit ourselves to events in O3 for self-consistency, as the injections we perform to evaluate selection effects (see Section \ref{sec:selection}) use O3a and O3b detector sensitivities.

We compute Eq.~\ref{likelihood} using the package \texttt{GWPopulation} \citep{Talbotgwpop}, which constructs a Monte Carlo approximation of this integral by reweighting samples from the single-event posteriors into the population model. We use the nested sampling package \texttt{Dynesty} \citep{Speagledynesty} to obtain hyperparameter samples from the posterior distribution. 

\subsection{Selection Effects}\label{sec:selection}
Since the sensitivity to an event is determined by the single-event parameters $\eta$, the observed population is biased toward events produced by the astrophysical population that are preferentially observable. 
To account for the bias arising from selection effects, we must compute the fraction of signals ($p_{\rm det}$) that will pass our detection threshold $\rho_{\rm th}$ by marginalizing over all possible signals and noise realizations $n$ (e.g., \cite{Vitale2022b, LIGO_O1})
\begin{equation}\label{pdet_analytic}
    p_{\rm det}(\Lambda) = \int d n \int d \eta p(\eta | \Lambda) p(n) \Theta(\rho - \rho_{\rm th}).
\end{equation}
Here $\Theta$ is the Heaviside step function.
In practice, we use Monte Carlo importance sampling with respect to some simulated fiducial reference population $\Lambda_0$ to estimate Eq.~\ref{pdet_analytic}. This method relies on injecting $N_{\rm inj}$ sources from this reference population into detector noise and determining which of these sources pass our detection threshold \citep{Farr19}. This is computed as
\begin{equation}\label{pdet}
    p_{\textrm{det}}(\Lambda) = \frac{1}{N_{\text{inj}}} \sum_{\eta \sim \eta_{\text{found}}} \frac{\pi(\eta|\Lambda)}{\pi(\eta|\Lambda_0)},
\end{equation}
where $\eta_{\text{found}}$ corresponds to the single-event parameters of the events from the injection set that pass detection threshold. 

For our sensitivity injection set, we simulate $\mathcal{O}(5 \times 10^7)$ sources and inject them into Gaussian noise corresponding to O3 detector sensitivity specified by the representative Power Spectral Densities in \citep{gwtc3, gwtc2}. This results in $\sim 900,000$ injections passing our detection threshold of network optimal SNR greater than 10, where the square of the network optimal SNR is defined as the quadrature sum of the SNRs in each detector. We choose this threshold to be a surrogate for the 1 year$^{-1}$ FAR threshold used for event selection. While this is not an exact mapping between the two detection statistics, the effects of spins on sensitivity is subdominant, so we expect that this approximation will not cause biases.

\subsection{Uncertainties in the Likelihood}

Since we approximate Eq.~\ref{likelihood} and Eq.~\ref{pdet} using Monte Carlo summation, there exists a resulting statistical uncertainty associated with the use of finite samples to obtain estimates for the value of the log likelihood \citep{Farr19, GolombTalbotbns, montecarlo, Essick22}. For each sample of $\Lambda$, we compute this associated uncertainty in the log likelihood. Considering the computed approximation of $\ln \mathcal{L}(\{d\}|\Lambda)$ to be a realization from a distribution that asymptotically tends to a Gaussian distribution, neglecting the contribution from $p_{\det}$, we compute the variance associated with the estimate as
\begin{equation}
    (\Delta \ln\mathcal{L}(\{d\}|\Lambda))^2 = \sum^N_i \left(\frac{\Delta \mathcal{L}_i}{\mathcal{L}_i}\right)^2,
\end{equation}
where $\mathcal{L}_i$ is the contribution to the likelihood from the $i$th event. We compute this fractional uncertainty as outlined in Appendix C of \cite{GolombTalbotbns}. 

We calculate the uncertainty from the Monte Carlo integration used to compute $p_{\rm det}$ in Eq.~\ref{pdet}, given an estimate of the mean $\mu$ as \citep{Farr19}
\begin{equation}
    (\Delta p_{\text{det}}(\Lambda))^2 = \frac{1}{N_{\text{inj}}^2} \sum_{\eta_{\text{found}}} \left(\frac{\pi(\eta|\Lambda)}{\pi(\eta|\Lambda_0)}\right)^2 - \frac{\mu^2}{N_{\textrm{inj}}}. 
\end{equation}

We can then calculate the total Monte Carlo uncertainty in the log likelihood by standard error propagation. An extended discussion of the calculation and the effects of this uncertainty will be the subject of future work. We note that considering $(\Delta \ln \mathcal{L})^2$ as the relevant quantity for avoiding biases due to the Monte Carlo approximation differs from what is suggested in \cite{Essick22} where the authors focus on the uncertainty in the difference between log-likelihood values.

While we do not enforce any threshold directly on $(\Delta \ln \mathcal{L})^2$, we retain this information for all points in the hyperposterior to investigate correlations between features in the population and uncertainty in the log likelihood (see Appendix \ref{appendix:injections}). 

\subsection{Uncertainty in the Evidence}

The evidence, or marginal likelihood, associated with a particular model is expressed simply as the expectation value of the likelihood conditioned on the population prior:

\begin{equation}\label{evidence}
    \mathcal{Z} = \int d\Lambda \mathcal{L}(\{d\}|\Lambda) \pi(\Lambda).
\end{equation}
Comparing this quantity for two different models allows one to compute a Bayes factor, which is commonly used as a discriminator between models based on their relative strength at describing the observed data.

Because \texttt{Dynesty} computes the evidence by iteratively summing over a finite number of weights, there exists a statistical uncertainty associated with the estimated evidence. \texttt{Dynesty} reports this uncertainty along with the computed evidence. 

Since there is also an uncertainty in the quantity $\ln\mathcal{L}((\{d\}|\Lambda)$used in computing the evidence, and the evidence is the average of a set of $\ln\mathcal{L}$ values, we take the contribution of this uncertainty to the total evidence to be the average uncertainty in $\ln\mathcal{L}((\{d\}|\Lambda)$ over the draws from $\pi(\Lambda)$. 

We take these two sources of uncertainty in the evidence to be independent and compute the total uncertainty in $\mathcal{Z}$ by propagation of errors. As a result, we obtain both the evidence and its uncertainty for each model of a fixed set of spline nodes. We note that all of the evidences for the spline models are consistent within their 1$\sigma$ uncertainties, and the \texttt{Default} model has a log Bayes Factor of $\approx -1.5$ with respect to the overlapping region of uncertainties in the evidences for the spline models.

\section{Results}\label{results}

\begin{figure}
    \centering
    \includegraphics[scale=0.6]{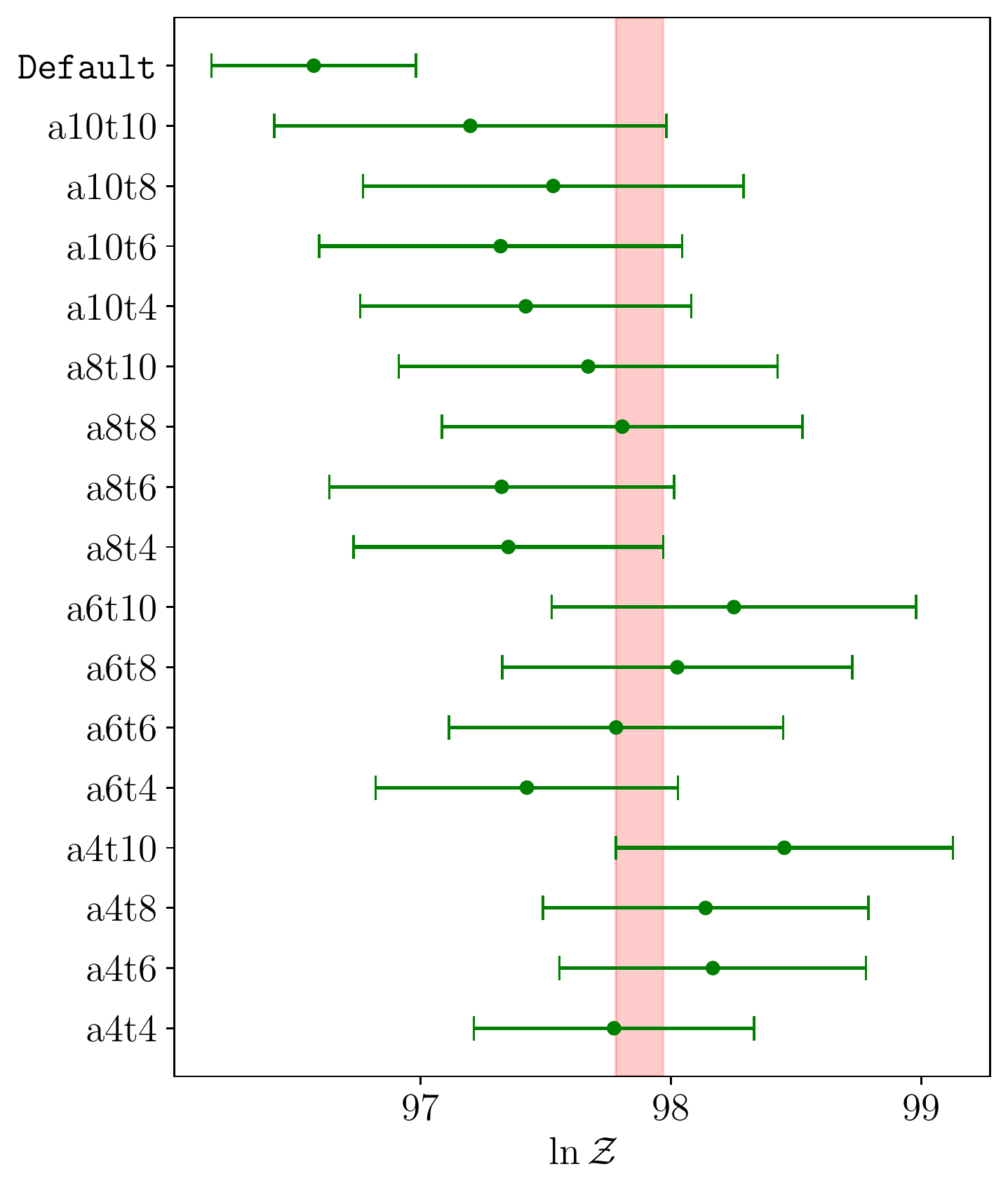}
    \caption{Comparison of evidences obtained from the different spline node combinations considered, as well as from the \texttt{Default} model. Uncertainties are computed by adding the average per-hyperparameter sample log likelihood uncertainty in quadrature with the uncertainty in the evidence as reported from \texttt{Dynesty}. The numbers after ``a'' and ``t'' are the number of nodes in the magnitude and tilt models, respectively. All evidences from the spline models are consistent with $\ln \mathcal{Z}$ in the red shaded region at 1$\sigma$.}
    \label{fig:evidences}
\end{figure}

In this section, we present the results from analyzing the population of BBH spin magnitude and tilts using spline models. We use the standard Gaussian prior on node amplitudes as described in Section~\ref{sec:splinepriors}, with nodes placed linearly within the domain of parameter space.
In Figure \ref{fig:evidences}, we show the evidences and their uncertainties for the 16 node combinations we consider in this work. All models give similar evidences, with no significant preferences considering their associated uncertainties. The red shaded region shows where all of the evidence estimates overlap within $1\sigma$. This indicates that adding more nodes does not tend to provide a better fit to the distribution and also does not over-fit it. We therefore cite the numbers in this section using the most flexible model, with 10 nodes for both the magnitude and tilt distributions. Unless otherwise noted, the plots of the spin magnitude (tilt) distributions assume 10 nodes in the tilt (magnitude) distribution.

As a general trend, we notice that the inferred 90\% region of parameter space exhibit oscillating peaks at the location of the spline nodes. As shown in Figure \ref{fig:priorsonly}, these oscillations appear for uninformative data. With the observations of BH spins being weakly informative, we see this effect from the strong influence the prior on the posterior distribution of the spline nodes.

\subsection{The distribution of spin magnitudes}

\begin{figure}
  \includegraphics[width=\linewidth]{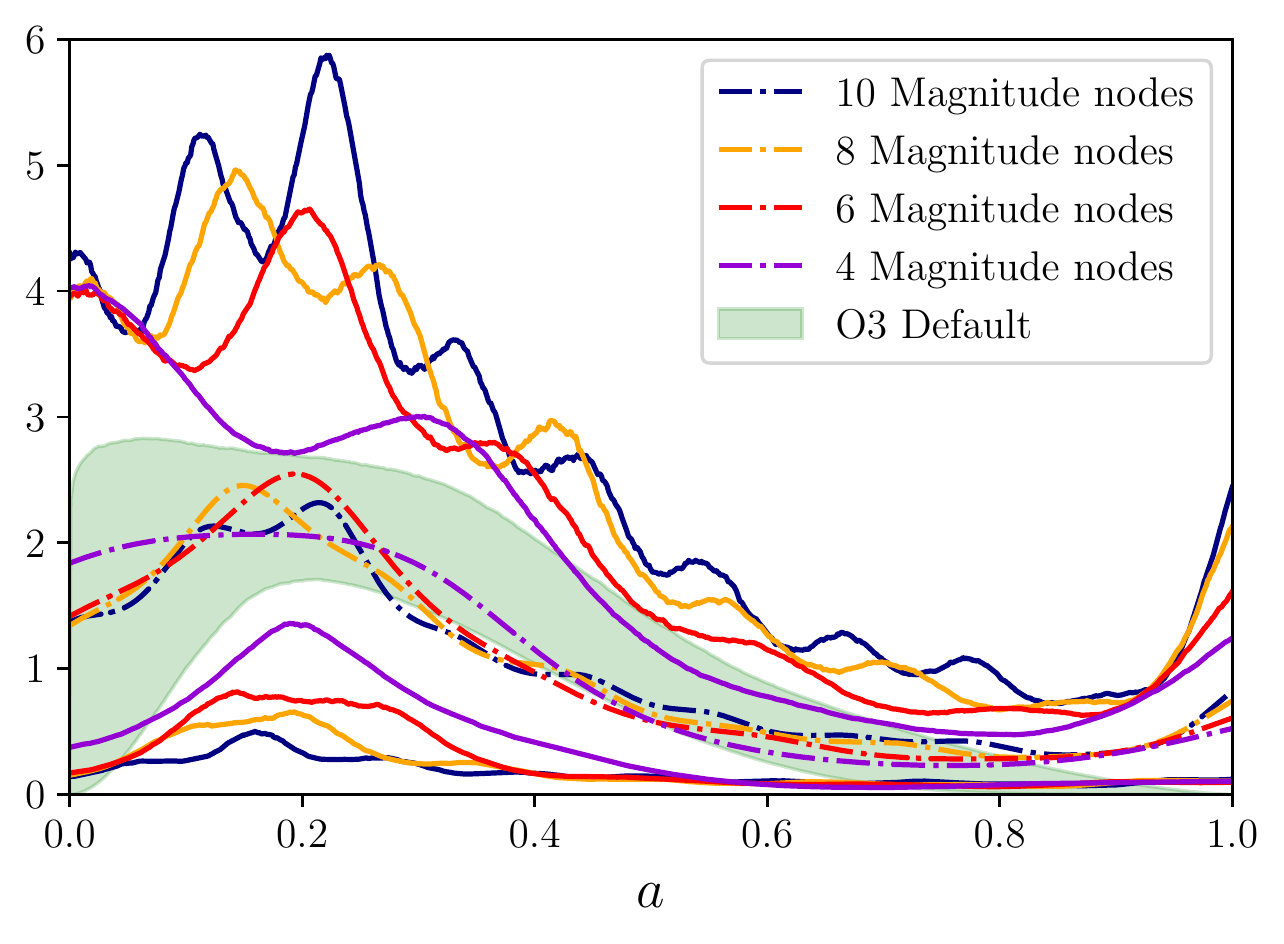}
  \caption{Distribution of spin magnitudes, with different numbers of nodes corresponding to different colors. All use 10 nodes in the tilt distribution. }
  \label{fig:spinmagnitudes}
\end{figure}

In Figure \ref{fig:spinmagnitudes} we show the inferred distribution of spin magnitudes for our four choices of node numbers in spin magnitude, assuming 10 nodes in spin tilt. Although each model involves different positions and numbers of spline nodes, we note that the uncertainties (solid lines) and the average line (dot-dashed lines) in the distribution are comparable between models. 

The 90\% credible interval of the distribution is relatively broad, making it difficult to discern obvious trends in spin magnitude. However, we note the general pattern of a preference for smaller spin magnitudes in the population and less support for higher spin magnitudes. Considering the model with 10 magnitude and 10 tilt nodes, we infer that ${77.1\%}^{+10.4\%}_{-14.8\%}$ of spin magnitudes are below $a=0.5$, and 50\% of spin magnitudes are below $a = {0.25}^{+0.16}_{-0.10}$ (all uncertainties in this work are reported at the 90\% symmetric credible levels unless otherwise stated).

While the models using fewer spline nodes tend to place increased support around $a=0.2$, the significance is substantially reduced as we add more spline nodes. When more spline nodes are added, the model becomes more flexible and more data are necessary to constrain the distribution. While this feature may be real, it is not confident enough to remain present as the flexibility of the model increases.

Comparing to \texttt{Default} model (the green shaded region) used in \cite{gwtc3pops}, we observe substantially more uncertainty in the inferred spin magnitude distributions using our spline models. By construction, the \texttt{Default} model requires $p(a = 0) = 0$, but this is ruled out from the spline model at the 90\% level. \footnote{Our spline model does not support $p = 0$ for any point in parameter space, as the spline is exponentiated, but the prior does allow for arbitrarily low values and we would therefore expect that the posterior for $p(a = 0)$ would approach 0 if the data truly favored the non-presence of zero-spin BHs.}
As a point of comparison, in Appendix \ref{appendix:prior} we show the distribution of spin magnitudes for different numbers of nodes, this time assuming 4 nodes in the tilt distribution. We find no significant differences from the distributions assuming 10 nodes in the tilt distribution which indicates that the number of nodes in the tilt distribution has a negligible effect on the inferred spin magnitude distribution.

\subsection{The distribution of spin tilts}

\begin{figure}
  \includegraphics[width=\linewidth]{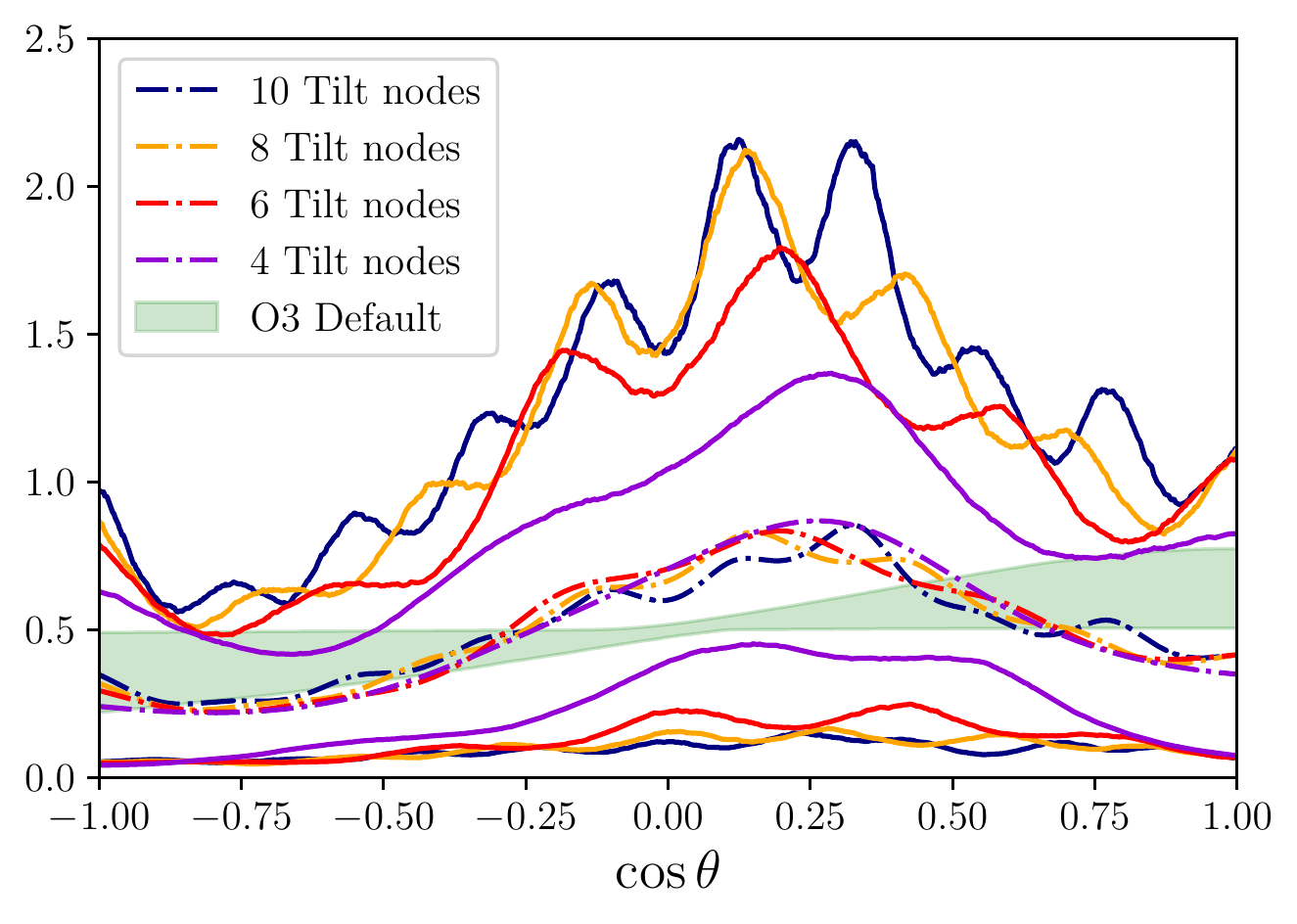}
  \caption{Distribution of spin tilts, with different numbers of nodes corresponding to different colors. All use 10 nodes in the spin magnitude distribution.}
  \label{fig:spintilts}
\end{figure}

In Figure \ref{fig:spintilts} we show the inferred distribution of spin tilts. Similar to the case with spin magnitudes, the uncertainties in the distribution are wide, but the average distributions for the different node combinations agree. In general, the distribution is consistent with being flat and featureless, but there is a slight trend for an increase in support for $-0.25 < \cos\theta < 0.75$. 

As demonstrated by comparing Figure \ref{fig:spintilts4} with \ref{fig:spintilts}, the inferred distribution of spin tilts is very similar when we model the spin magnitude distribution with 4 nodes. We confirm this invariance for all sets of magnitude nodes tested, suggesting that the number of spin magnitude nodes does not meaningfully affect the recovered spin tilt distribution. 

We infer that ${38.6\%}^{+17.3\%}_{-15.6\%}$ of spin tilts are below $\cos\theta=0$, and 50\% of spin tilts are below $\cos\theta = {0.15}^{+0.22}_{-0.22}$. Notably there is no trend for an increase in support for $\cos\theta = 1$ as would be predicted by a preferentially aligned-spin population (see Section \ref{discussion}).

\subsection{The distribution of $\chi_{\rm eff}$}

\begin{figure}
  \includegraphics[width=\linewidth]{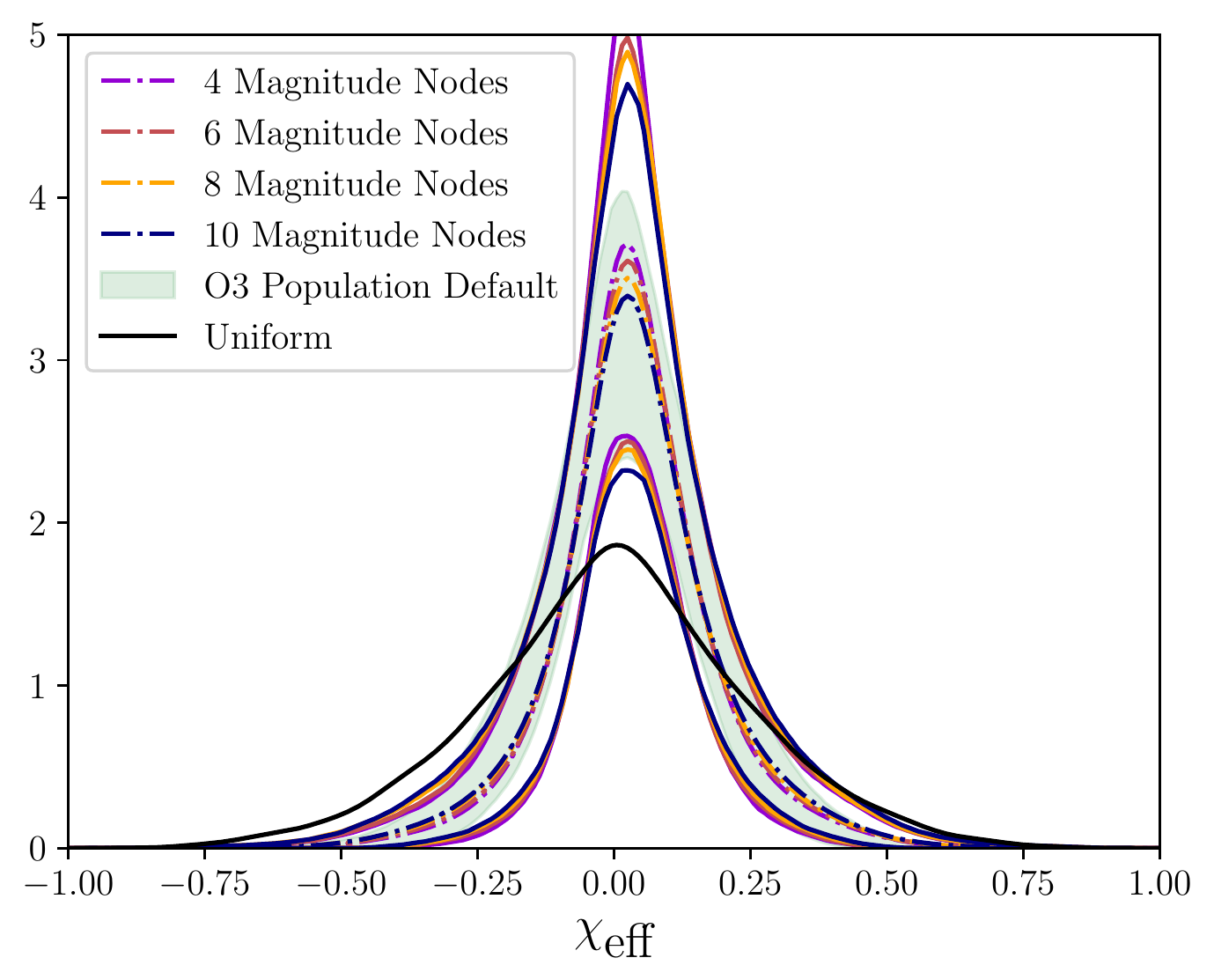}
  \includegraphics[width=\linewidth]{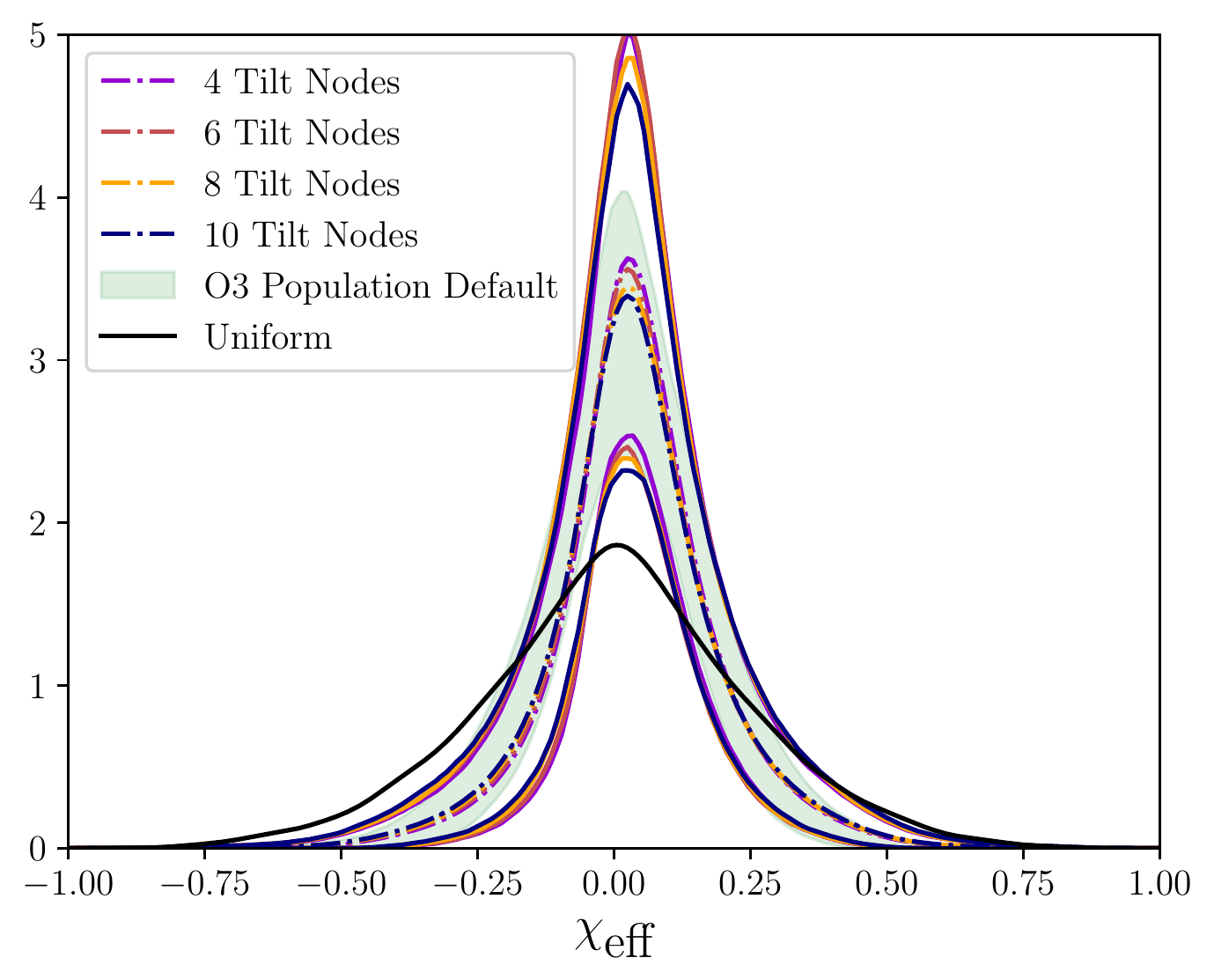}
  \caption{Distribution of effective inspiral spin parameter as recovered from the distributions in Figure \ref{fig:spintilts} and Figure \ref{fig:spinmagnitudes}}
  \label{fig:chieff}
\end{figure}

As an alternative to modeling the component spins, it is common to consider instead the total spin contribution aligned with the orbital angular momentum, the so-called ``effective'' aligned spin parameter. This term is parameterized as
\begin{equation}
    \chi_{\rm eff} = \frac{a_1\cos\theta_1 + q a_2\cos\theta_2}{1 + q}.
\end{equation}

While we do not directly model the distribution of $\chi_{\text{eff}}$ in this work, we can use the inferred distributions of $a$, $\cos(\theta)$, and $q$ to reconstruct a distribution for $\chi_{\rm eff}$ (c.f. \cite{gwtc2pops, Galaudage21}). 

Figure \ref{fig:chieff} shows this inferred distribution of $\chi_{\rm eff}$ as we vary the number of tilt and magnitude nodes. As a point of comparison, we show the corresponding reconstruction of $\chi_{\rm eff}$ when the distributions of component spin magnitudes and tilts are inferred using the \texttt{Default} model with same catalog of events. In addition, we also plot the $\chi_{\rm eff}$ distribution recovered from uniformly sampling in $a$ and $\cos(\theta)$ (solid black curve), assuming mass ratios drawn from a power law with an index of 2 ($q$ consistent with the results in \cite{gwtc3pops}) . The $\chi_{\rm eff}$ distribution inferred from the spline model agrees well with the \texttt{Default} reconstruction, but is a noticeably narrower distribution than a uniform spin magnitude and tilt distribution would result in.
Using the model with ten magnitude and tilt nodes each, we infer that ${38.7\%}^{+12.8\%}_{-11.5\%}$ of BBH systems have $\chi_{\rm eff} < 0$.

\subsection{The distribution of $\chi_{\rm p}$}

\begin{figure}
  \includegraphics[width=\linewidth]{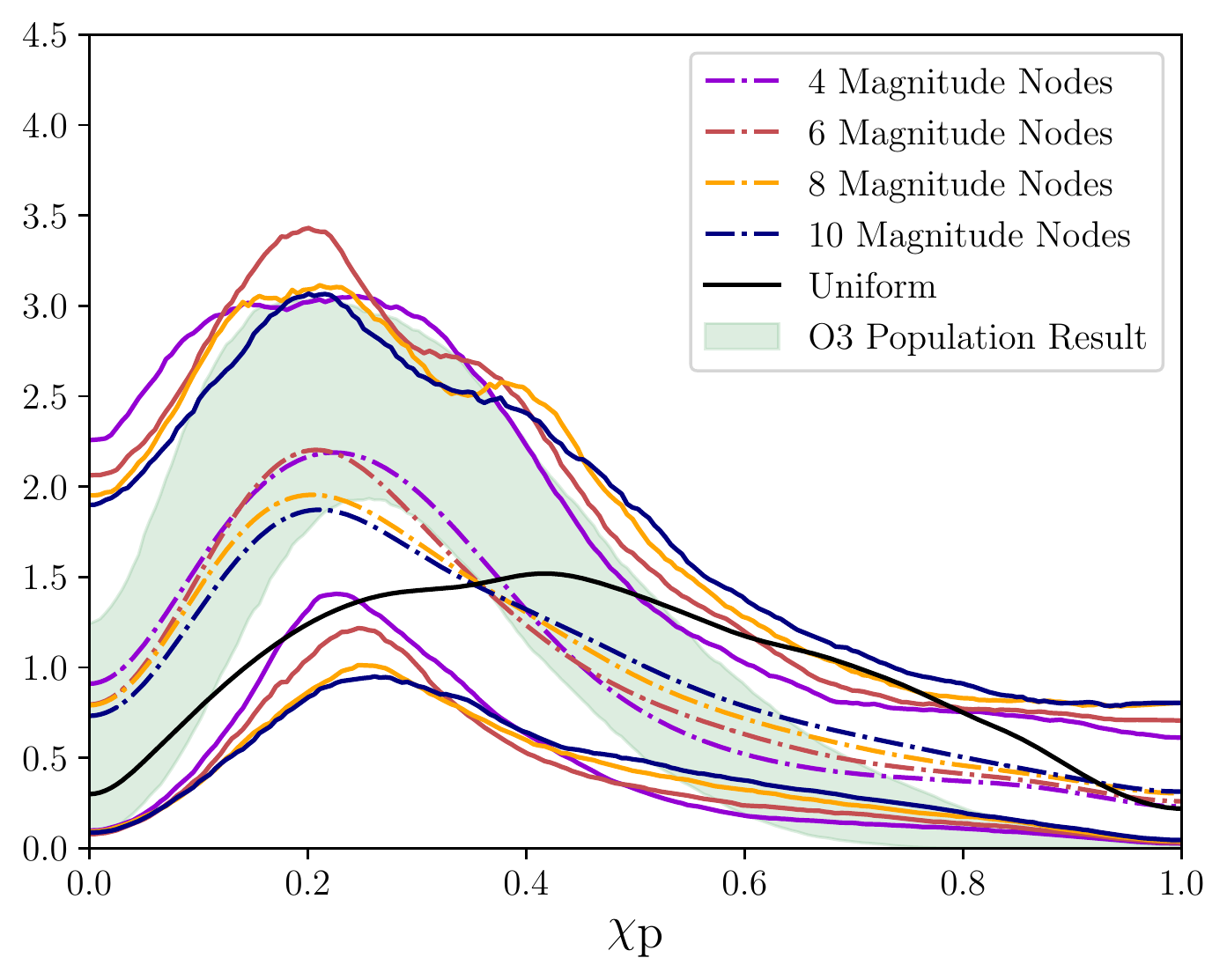}
  \includegraphics[width=\linewidth]{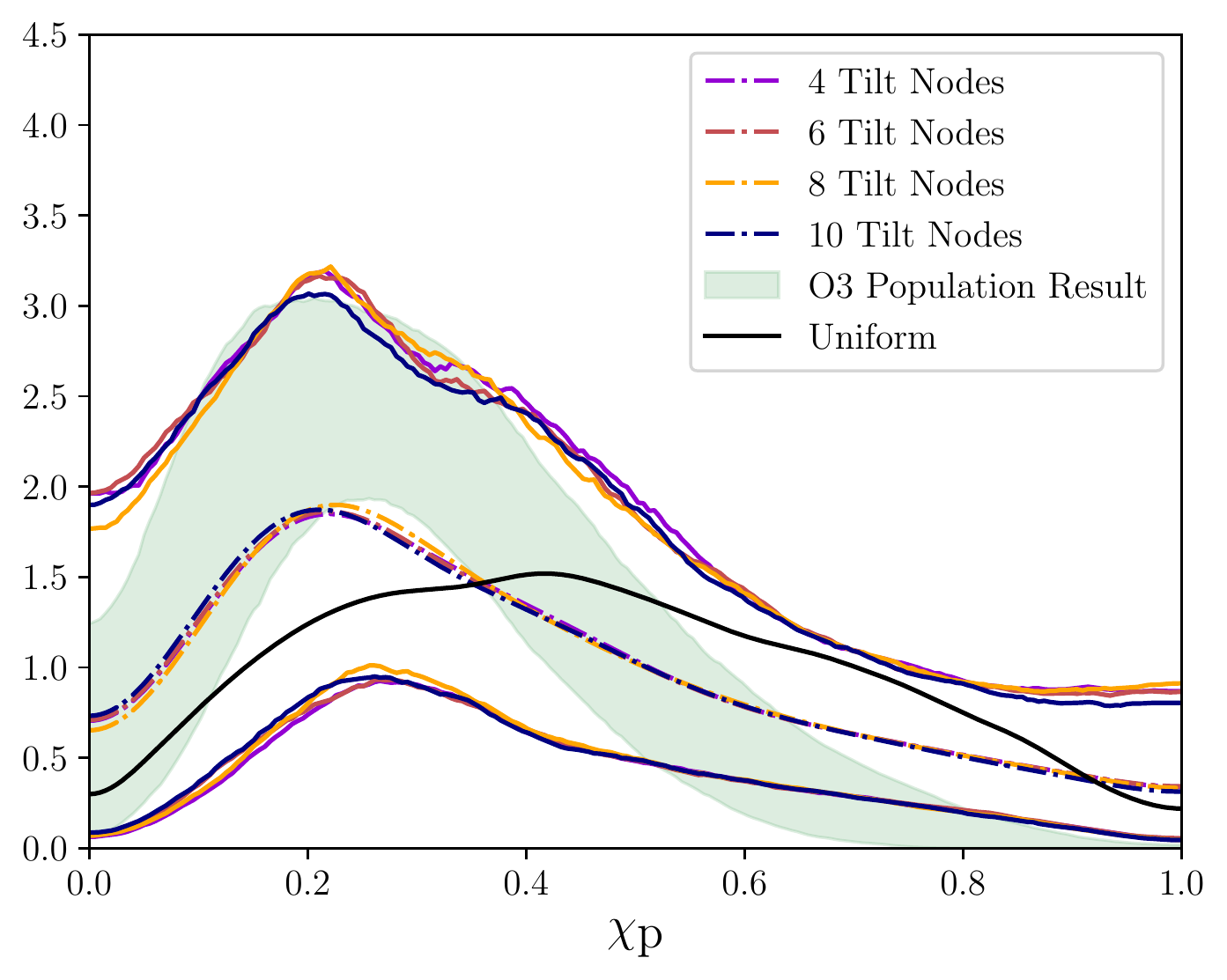}
  \caption{Distribution of precession spin parameter as recovered from the distributions in Figure \ref{fig:spintilts}.}
  \label{fig:chip}
\end{figure}

Another ``effective'' spin parameter commonly modeled in the gravitational wave literature is the effective precessing spin parameter, $\chi_{\rm p}$, which quantifies the amount of in-plane spin present in a BBH merger \citep{Schmidt15}. Here,
\begin{equation}
    \chi_{\rm p} = \max \left[\ a_1 \sin\theta_1, \left(\frac{3 + 4q}{4 + 3q}\right) q a_2 \sin\theta_2\right]\ .
\end{equation}
Similar to the previous subsection, we can reconstruct the distribution of $\chi_{\rm p}$ using the spline models of $a$, $\cos(\theta)$, as well as our inference on the population of $q$.

Figure \ref{fig:chip} shows the inferred distribution of $\chi_{\rm p}$ as a function of different magnitude and tilt nodes, respectively. We also show the inferred distribution recovered from the \texttt{Default} model \citep{gwtc2pops, gwtc3pops, Talbotspin, Wysocki19} analysis using the same event list, as well as the distribution corresponding to uniform distributions in $a$ and $\cos\theta$. The inferred distribution of $\chi_{\rm p}$ is consistent with an isotropic distribution, and shows agreement with the $\chi_{\rm p}$ distribution reconstructed from the \texttt{Default} model. An exception to this agreement is the slightly increased support at high $\chi_{\rm p}$ that is not present in the \texttt{Default} reconstruction. The higher support for large $a$ and $\cos(\theta) \approx 0$ in the spline model relative to the \texttt{Default} model explains to this increased support at high $\chi_{\rm p}$. Similarly, our result allows for more support at high $\chi_{\rm p}$ relative to what is presented in \cite{gwtc3pops}, in which it is assumed that the $\chi_{\rm eff}$ and $\chi_{\rm p}$ distributions follow a multivariate normal distribution \citep{Miller2020}. In particular, Figure 16 in \cite{gwtc3pops} shows vanishing support for $\chi_{\rm p} > 0.4$ in the population, whereas we find some support in this region is included at the 90\% credibility. 
We note that varying the number of nodes in spin magnitude has the largest impact on the averaged recovered $\chi_{\rm p}$ distribution (dash-dotted curves) indicating that this measurement depends on our choice of prior for the spin magnitude distribution.

\section{Discussion}\label{discussion}

Most previous analyses of the astrophysical distribution of merging binary black hole systems have focused on fitting parametric phenomenological models strongly constrained by the functional form of the model to the observed data (e.g.,~\cite{gwtc3pops} and references therein) or directly compared with detailed simulations (e.g.,~\cite{Zevin2021, Wong2021}).
However, more data-driven methods have been employed to infer the binary black hole mass~\cite{Mandel2017, Tiwari2021, Rinaldi2022, Edelman} and spin distributions~\cite{Tiwari2021}.
    
In this work, we use cubic splines (a model previously used to fit the black hole mass distribution \cite{Edelman}) to fit the astrophysical spin magnitude and spin tilt distributions of black holes as inferred from LIGO-Virgo observations. In doing so, we limit the influence of prior modelling assumptions on the inferred distribution and present a more data-driven result. While the uncertainties in the inferred distributions are large, we are able to interpret trends as they relate to astrophysical mechanisms of BBH formation. 

Models of stellar physics suggest that angular momentum transport out of the core of a collapsing star is highly efficient, indicating that first generation stellar BHs should primarily have negligible spin \citep{FullerMa15}. Based on this, some models tend to favor nonspinning BHs when born in isolated environments and not susceptible to tidal spin-up. 

Motivated by the work in \cite{FullerMa15, Belczynski20}, the authors of \cite{Galaudage21, Callister22, gwtc3pops, Tong2022} search for contributions from a nonspinning subpopulation of BHs in the distributions of $\chi_{\rm eff}$ and spin magnitude. Some of this previous work has found support for $a = 0$ when using low-dimensional parametric spin distribution models that allow for support at that point \citep{Talbot20, Wang22, Galaudage21, Kimball20, Kimball21}.
The results of such inference are strongly model dependent, with the preference for the presence of a non-spinning component depending on the morphology of that component. 
We do not confidently recover such a feature as demonstrated in Figure \ref{fig:spinmagnitudes}. Given the width of the 90\% credible interval at low spin magnitudes, we are unable to rule out the presence of this feature. This is consistent with what was found in \citep{Tong2022, Callister22, Mould2022}, in which the authors find that there is insufficient data to resolve such a non-spinning subpopulation when employing a strongly parametric spin model with a spike at $a = 0$. 

On the other hand, the merger of equal mass, nonspinning BBH systems are expected to result in a remnant BH with  $a \sim 0.7$. As a result, population simulations predict that hierarchical mergers resulting from products of nonspinning first-generation mergers will leave a signature of a subpopulation of BH spins peaked around $a \sim 0.7$ with tails extending from $a \sim 0.5$ to $\sim 0.9$ \citep{Fishbach17}. Referring back to Figure \ref{fig:spinmagnitudes}, we do not see evidence of an obvious subpopulation in this high-spin region of interest, but rather some preference for low-spin magnitudes, possibly indicating that hierarchical mergers are not providing the dominant formation mechanism for the observed BBHs. The lack of support for a relatively high spin subpopulation is consistent with the conclusions drawn in \cite{gwtc3pops, Callister22}. Assuming that BBHs from hierarchical mergers all have $a>0.5$, we infer that no more than ${23\%}^{+14\%}_{-11\%}$ of the astrophysical population of merging BBHs form through a hierarchical merger channel.

A notable feature in our analysis is the increased uncertainty in the spin magnitude distribution as compared to that from the Beta distribution in spin magnitude. The motivation for using a Beta distribution in \cite{Wysocki19, gwtc2pops, gwtc3pops} is not physical but is statistical: the Beta distribution only has support in the interval [0, 1] and offers a flexible, parametric fit for the mean and variance of a distribution and has an analytic form. The spline model introduced in this work offers more flexibility than the Beta distribution, so lacking a physical motivation for the Beta distribution, we expect that the uncertainties in the spin magnitude distribution obtained in this work are more appropriate than those obtained from the \texttt{Default} model. Furthermore, the Beta distribution used in \cite{gwtc2pops, gwtc3pops} cannot model structures such as increased support for nonspinning BHs or a secondary peak at high spin, making it a suboptimal model for the astrophysical spin magnitude distribution in the presence of a nonspinning subpopulation. Comparing to our data-driven approach, we therefore conclude that the resulting spin distribution presented in \cite{gwtc3pops} is partially model-driven rather than data-driven.

When dynamical encounters take place within dense environments such as globular clusters, it is likely that some of the remnant BHs are retained in the cluster and merge in a subsequent dynamical encounter, contributing to the hierarchical merger population. The authors of \cite{Fishbach22} find that, for a broad range of populations considered, 16\% of mergers in the hierarchical merger population have $\chi_{\rm eff} < -0.3$. Using our inferred $\chi_{\rm eff}$ distribution we infer $2.1\%^{+3.9\%}_{-1.5\%}$ of BBH mergers have $\chi_{\rm eff} < -0.3$. Using this interpretation from the $\chi_{\rm eff}$ distribution, we place a conservative upper limit on the contribution of hierarchical mergers to the BBH merger population of ${13\%}^{+24\%}_{-9\%}$ which agrees with the one obtained when using just the spin magnitude information. This limit broadly agrees with the upper limit of 26\% for the fraction of hierarchical mergers presented in \citep{Fishbach22}, in which the authors use low-dimensional parametric models to infer the $\chi_{\rm eff}$ distribution.
This is also consistent with the results of~\cite{Kimball21} who found that depending on the escape velocity of the hierarchical merger environment up to $\approx10\%$ of merging black holes may come from hierarchical mergers.

Mergers of BBH systems which have spins that are isotropic in orientation, as is expected from dynamical formation scenarios, implies a distribution of $\chi_{\rm eff}$ symmetric about 0 (see the black line in Figure \ref{fig:chieff}). This prediction comes from the idea that during a dynamical capture, there is no reason to expect that the two BHs should have correlated spin directions when they randomly encounter each other. In contrast, spins of BBH systems forming from common evolution are expected to remain primarily aligned with the orbital angular momentum, resulting in exclusively positive values for $\chi_{\rm eff}$ from this population. While the distribution of $\chi_{\rm eff}$ we recover appears symmetric, it is centered at $\chi_{\rm eff} = {0.033}^{+0.034}_{-0.038}$, favoring a positive central location but consistent with being centered at $\chi_{\rm eff} = 0$ at the 90\% level. This constraint is similar to that obtained by \cite{Edelman22}, using basis splines to model the component spin distributions. This result, coupled with the result that ${38.7\%}^{+12.8\%}_{-11.5\%}$ of events have $\chi_{\rm eff} < 0$, presents the possibility that dynamical encounters are a significant contribution to the formation mechanisms of BBH merger systems. This is in contrast to the results reported in \cite{gwtc3pops} that the $\chi_{\rm eff}$ distribution is centered at 0.06 and rules out being centered at 0 at the 90\% level; this result is obtained by modeling the $\chi_{\rm eff}$ distribution as a Gaussian with the mean and standard deviation as free parameters \citep{Miller2020, gwtc3pops}. While we use a largely identical event list in our analyses, the modeling assumptions for the $\chi_{\rm eff}$ distribution made in \cite{gwtc3pops} probably explain some of the differences in our results. While \cite{Roulet, Callister22} also find a $\chi_{\rm eff}$ distribution consistent with being centered at zero, \cite{Roulet} does not find any support for $\chi_{\rm eff} < 0$; such a difference may also be due to different modeling choices for the population of $\chi_{\rm eff}$.

We see increased uncertainty in $\cos\theta$ with respect to those obtained from the \texttt{Default} model from \citep{gwtc3pops}. While the tilt distribution of the \texttt{Default} model is astrophysically-motivated, it is incapable of capturing any possible substructure that may be present at locations other than $\cos\theta = 1$, as it is modeled by a monotonic function. Given the additional flexibility of the spline model, we notice a trend in the average line of the $\cos(\theta)$ distribution toward increased support for $-0.25 < \cos\theta < 0.5$. This trend is of low significance given the uncertainties surrounding it in the inferred $\cos\theta$ parameter space, but may indicate a nontrivial contribution from BBHs with in-plane spins to the astrophysical population. This trend is consistent with what is found in \citep{Vitale2022, Edelman22}, in which the authors use more flexible models to infer the $\cos\theta$ distribution. Our result that  ${38.6\%}^{+17.3\%}_{-15.6\%}$ of BHs exhibit negative spin tilts is broadly consistent with previous studies that indicate the need for negative alignment in the astrophysical population. The presence of support for $\cos\theta < 0$ in the population was reported in \cite{gwtc2pops} and confirmed in subsequent studies (e.g. \cite{Callister22, gwtc3pops}).

It is generally considered unlikely for BBH systems formed under common/isolated evolution scenarios to exhibit spin-orbit misalignment, as any such misalignment in these systems is expected to be corrected by mass transfer and tidal effects \citep{Farr17, Kalogera2000}.  
Lack of confidently-increased support for $\cos\theta = 1$ indicates that aligned-spin BBH systems do not contribute a statistically resolvable subpopulation of mergers. A possible explanation for this is a comparable or more significant contribution of BBH mergers from dynamical encounters in dense environments to the inferred astrophysical population of BBH mergers, as this would manifest as a more isotropic distribution in tilts.

The work presented in this paper motivates the need for more data-driven models for inferring the BBH spin distribution from GW sources, as there may be features of astrophysical importance that cannot be captured by currently-used parametric models. While we cannot confidently discern many trends in the spin magnitude and tilt distributions, we can place constraints on the support in different parts of spin parameter space by substantially relaxing modeling assumptions. We also show that the uncertainties in these distributions as inferred using strongly parametric models are artificially small if such a model does not well-describe the underlying distribution. Using our more flexible model, we find that substantially increased uncertainties are a necessary cost to being able to model arbitrary features in the spin distribution, given current GW data. Data collected from events in future observing runs may help resolve such features which may exist in the spin distribution, as well as motivate a better choice of priors to use on these data-driven models.

\begin{acknowledgments}

We thank Ethan Payne, Salvatore Vitale, Derek Davis, and Sylvia Biscoveanu for helpful discussions regarding this work.
We are grateful to Bruce Edelman and Ben Farr for fruitful discussions of spline modeling. We additionally thank Chase Kimball for reviewing this manuscript.
JG acknowledges funding from NSF grants 2207758 and PHY-1764464.
CT is supported by an MKI Kavli Fellowship. This material is based upon work supported by NSF's LIGO Laboratory which is a major facility fully funded by the National Science Foundation.
This work used computational resources provided by the Caltech LIGO Lab and supported by NSF grants PHY-0757058 and PHY-0823459.
\end{acknowledgments}

\appendix

\section{Efficient evaluation of the spline model}\label{appendix:eval}

Our model requires evaluating a different spline model at many values during each likelihood evaluation.
This process can be divided into three stages: constructing the spline model, identifying where each of evaluation points lies in relation to the nodes, and evaluating the appropriate piece in the spline.
The first stage must be performed at every iteration but does not depend on the number of points that the spline will be evaluated at.
The second stage is independent of the value of the spline nodes but must be performed for each of the evaluation points.
For a uniform spacing of spline nodes, this can be efficiently evaluated, however, for a generic spline this can be computationally intensive.
At the third stage, we simply combine the results of the two previous stages.
This can be trivially parallelized using a graphics processing unit (GPU).

For our use case, the locations at which the splines are evaluated and the node points are the same at every iteration.
We can therefore cache the result of the second stage.
We find that for our application the caching method accelerates the evaluation of the model by a factor of $\gtrsim 100$.
Our implementation {\tt cached\_interpolate} is available via {\tt pypi} and {\tt conda-forge}.

\section{Comparison between injection sets}\label{appendix:injections}

\begin{figure}
  \includegraphics[width=\linewidth]{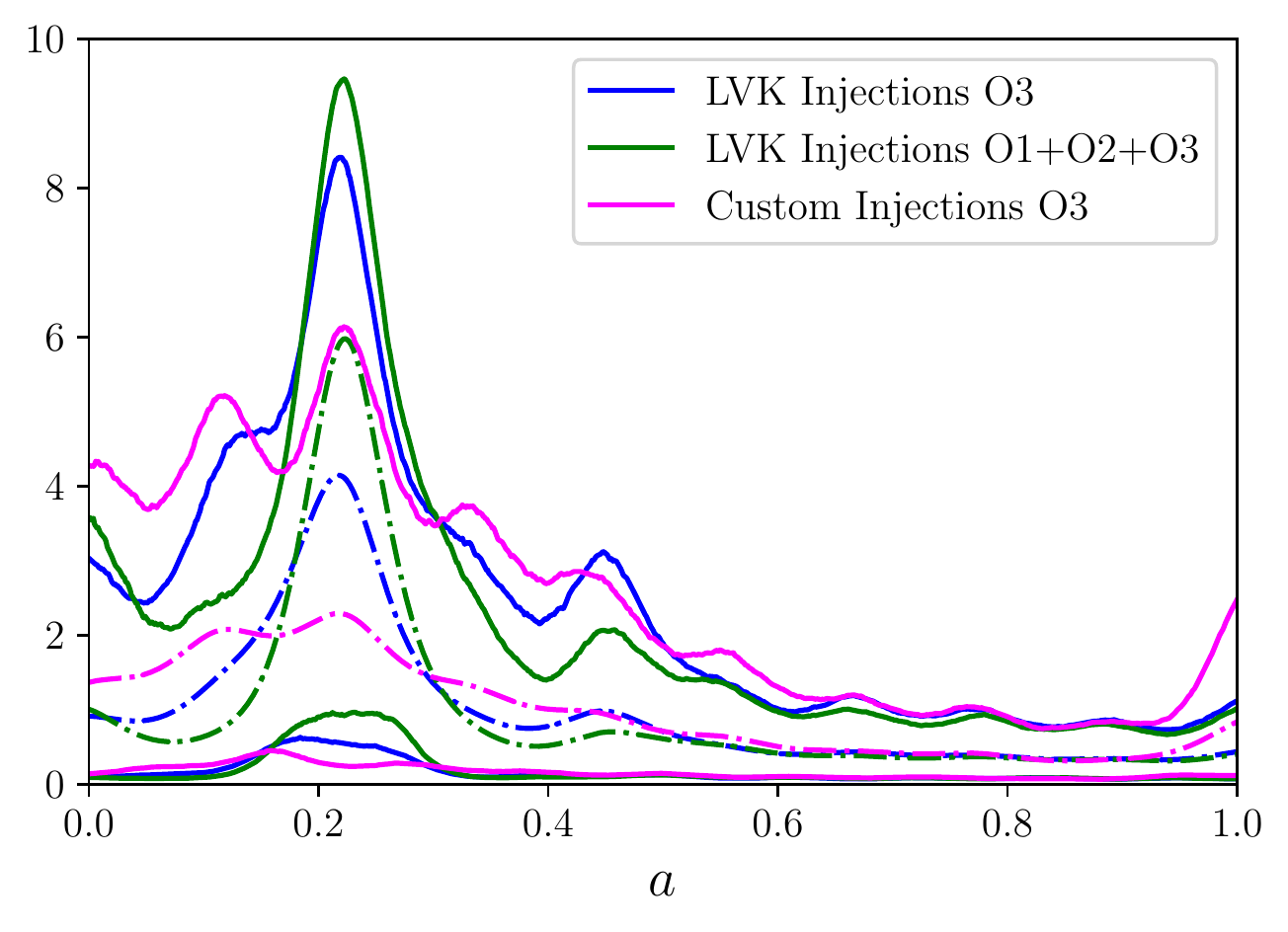}
  \caption{Distribution of spin magnitude using 10 nodes. Different colors correspond to different sensitivity injection sets. Injection sets use events from their corresponding observing runs. Note the peak at around $a=0.2$ is most pronounced with using O1+O2+O3 set and least with our custom O3 set.}
  \label{fig:compare_injections_tilt}
\end{figure}

\begin{figure}
  \includegraphics[width=\linewidth]{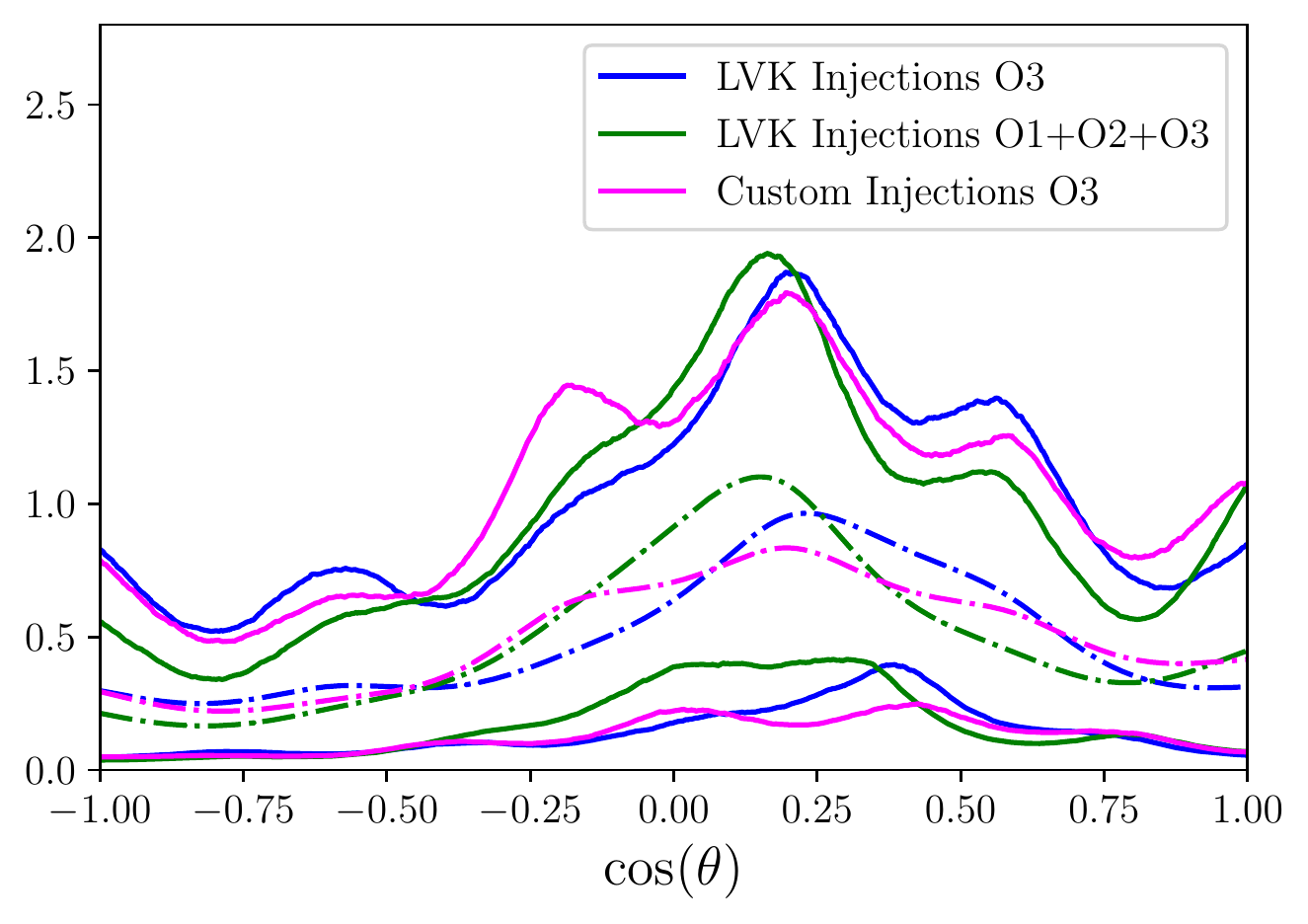}
  \caption{Distribution of spin tilt using 6 nodes. Different colors correspond to different sensitivity injection sets. Injection sets use events from their corresponding observing runs.}
  \label{fig:compare_injections_tilt}
\end{figure}

\begin{figure}
  \includegraphics[width=\linewidth]{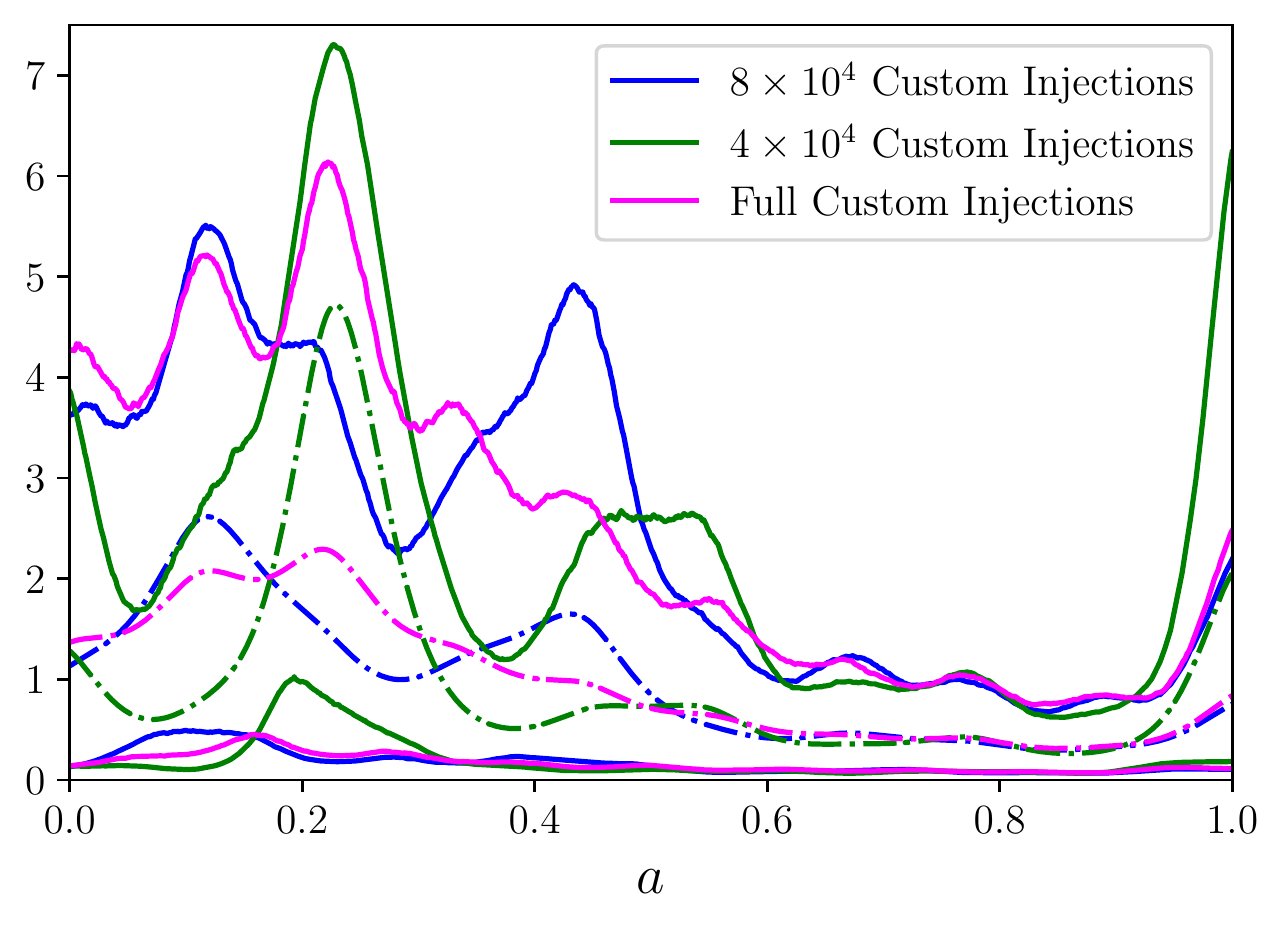}
  \caption{Distribution of spin tilt using 6 nodes. Different colors correspond to a different number of found injections: the 900,000 from the custom injection set, $\sim$ 80,000, and $\sim$ 40,000, where the latter two are close to the number of found injections in the LVK O3-only and O1+O2+O3 injection sets, respectively.}
  \label{fig:downsampled_a}
\end{figure}

\begin{figure}
  \includegraphics[width=\linewidth]{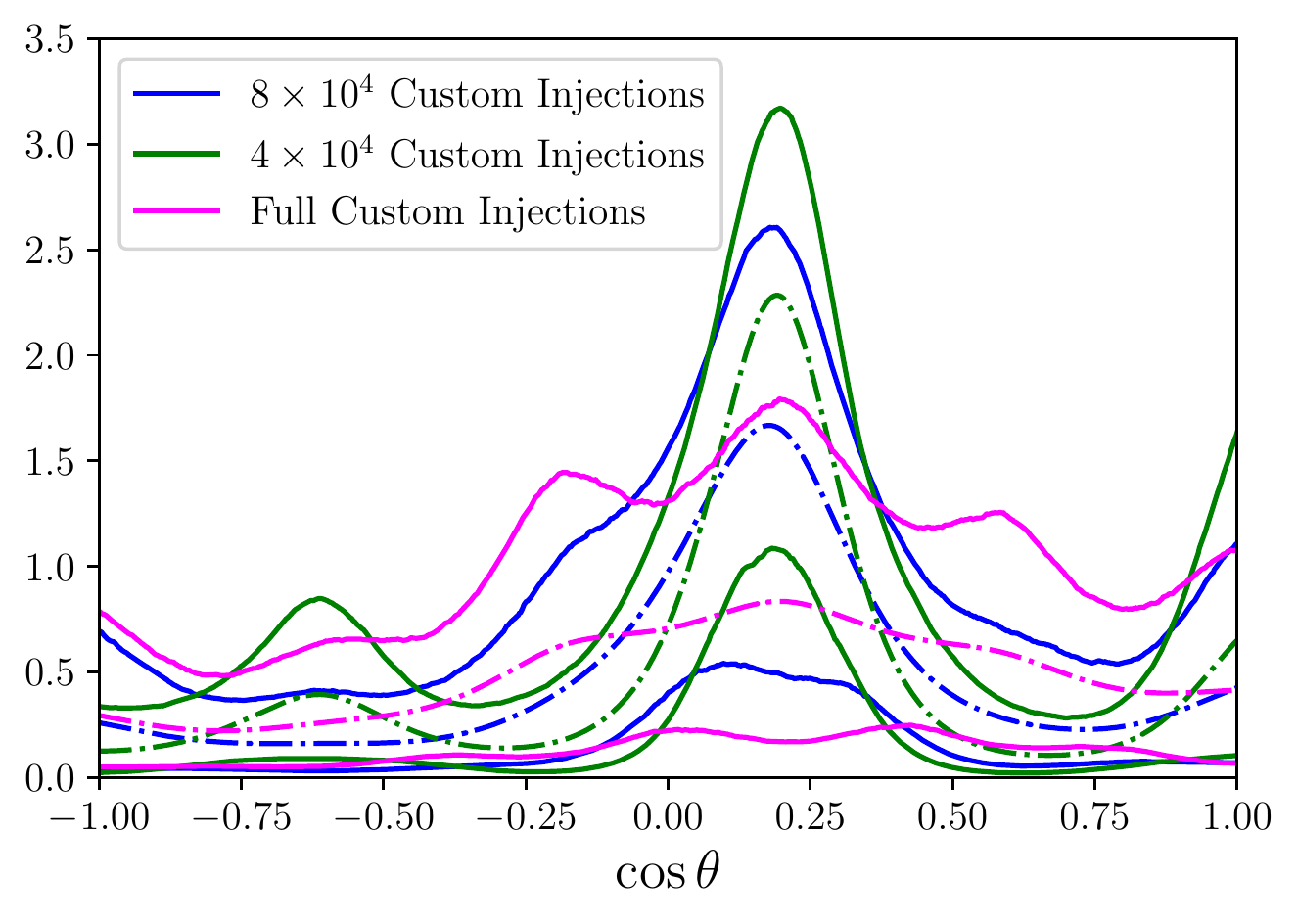}
  \caption{Distribution of spin tilt using 6 nodes. Different colors correspond to a different number of found injections: the 900,000 from the custom injection set, $\sim$ 80,000, and $\sim$ 40,000, where the latter two are close to the number of found injections in the LVK O3-only and O1+O2+O3 injection sets, respectively.
  }
  \label{fig:downsampled_tilt}
\end{figure}

Sensitivity estimates for Advanced LIGO and Virgo were released along with \cite{gwtc3, gwtc3pops} for the first three observing runs \citep{lvko1o2o3injections, o3lvkinjections, gwtc3}. These sensitivity estimates consist of injections of simulated sources into detector noise, along with the SNRs and FARs of these injections as reported by the detection pipelines used in the LIGO-Virgo observing runs. In this section, we compare the use these injections to compute the $p_{\rm det}$ term as written in Eq. \ref{pdet} to the use of our own injections to compute the same term.

Using a set of injections, we include those which pass a detection threshold in the summation over the found injections. For the injections provided in \cite{gwtc3}, we use a threshold of FAR $<$ 1 yr$^{-1}$, consistent with the choice made in \cite{gwtc3pops}. We do not run the detection pipelines to assign a FAR to each of the injections from our custom injection set in this paper, so we threshold these on an optimal SNR $>$ 10. Since importance sampling Monte Carlo integration relies on drawing enough samples from the fiducial distribution that cover the support of the target distribution, using Eq. \ref{pdet} as a reliable estimator for Eq. \ref{pdet_analytic} requires a suitable number of ``found'' injections to get a well-converged estimate (i.e. see \citep{Farr19}). With too few samples being used to compute the Monte Carlo approximation, the variance of our estimator is large and the resulting estimate may be a poor approximation of the true log likelihood. The statistical uncertainty in the log likelihood estimates at each point in parameter space can cause a systematic bias to appear in the resulting posterior distribution for the population.

\begin{figure}
  \includegraphics[width=\linewidth]{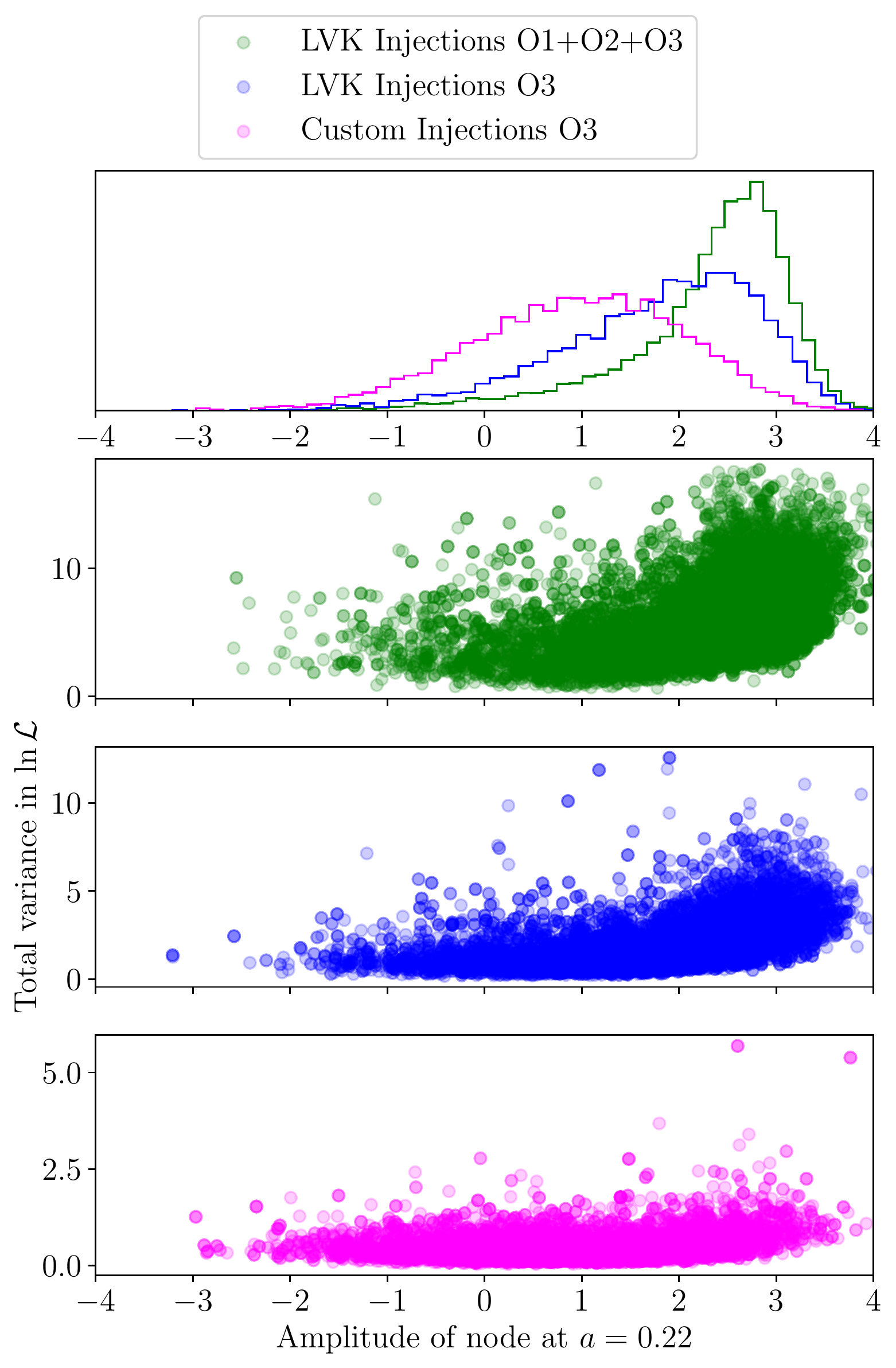}
  \caption{Posterior samples for the amplitude of the spline node at $a = 0.22$ and their associated variances in the log likelihood. Note the positive correlation between spline amplitude and variance. The results obtained using the injection set from all three observing runs, which has the least number of found injections, exhibits the highest variance in the log likelihood.}
  \label{fig:amplitude_unc}
\end{figure}

The injection sets provided in \cite{gwtc3, gwtc3pops} for the combined O1+O2+O3 sensitivity and O3 sensitivity respectively contain 41,972 and 81,117 simulated events which pass our detection threshold. As expected, the use of more samples reduces the uncertainty in the computed $p_{\rm det}$ estimate, and is reflected by the distribution of variances in $\ln \mathcal{L}$, as shown in Figure \ref{fig:amplitude_unc}. We drastically reduce the variances in the log likelihood estimates by using our own injection set which contains 911,386 injections passing our detection threshold.

In order to validate that the number of samples used to compute $p_{\rm det}$ is a cause of systematic bias in the inferred population (as opposed to the difference in detection statistic used for the threshold), we repeat the above spin distribution inference but using injection sets that have been downsampled to have $\sim 40,000$ and $\sim 80,000$ found injections. In Figures \ref{fig:downsampled_a} and \ref{fig:downsampled_tilt} we note recovery of strongly peaked features in the spin magnitude and tilt distributions, respectively. The significance of these features becomes drastically reduced as the number of found injections increases, indicating that a lower number of effective samples used to compute Equation \ref{pdet} can lead to biases that propagate into spurious features in the spin distribution. We therefore infer that a sufficient number of injections is necessary to recover an unbiased spin distribution using our spline model implementation, motivating our use of the custom injection set with substantially more found injections than what was released in \cite{o3lvkinjections}.

We see in Figure \ref{fig:amplitude_unc} that the amplitude of the node at $a = 0.22$ is correlated with higher statistical variance; as the amplitude of this node increases, the uncertainty in the log likelihood increases as well, making the log likelihoods computed in this part of parameter space less trustworthy. As we decrease the variance by using injection sets with higher $N_{\rm inj}$, the uncertainty in the log likelihood estimates decreases. With better estimates of the log likelihood, the support for the high amplitude of the node at $a = 0.22$ decreases, indicating that this feature in the spin magnitude distribution may be an artifact of poorly-converged Monte Carlo integrals. If this peak were a true feature in the astrophysical population, we would expect the inferred distribution computed with the custom injections would maintain support for high amplitude at this node, along with reduced uncertainty. We confirm that this uncertainty is associated with the selection function rather than associated with reweighting posterior samples in the population model by comparing the contributions of the uncertainties in both these Monte Carlo summations to the total propagated uncertainty in the log likelihood; for the models tested in this paper, we consistently find that the uncertainty associated with the contribution from Eq~\ref{pdet} dominates.

We note that we have noticed several other examples of similar behavior in our analyses, notably manifesting as spurious peaks in the spin distribution. This demonstrates the need for sufficient coverage of injections when using importance sampling to compute sensitivity estimates especially when evaluating a population distribution that can model narrow peaks.

\begin{figure}
  \includegraphics[width=\linewidth]{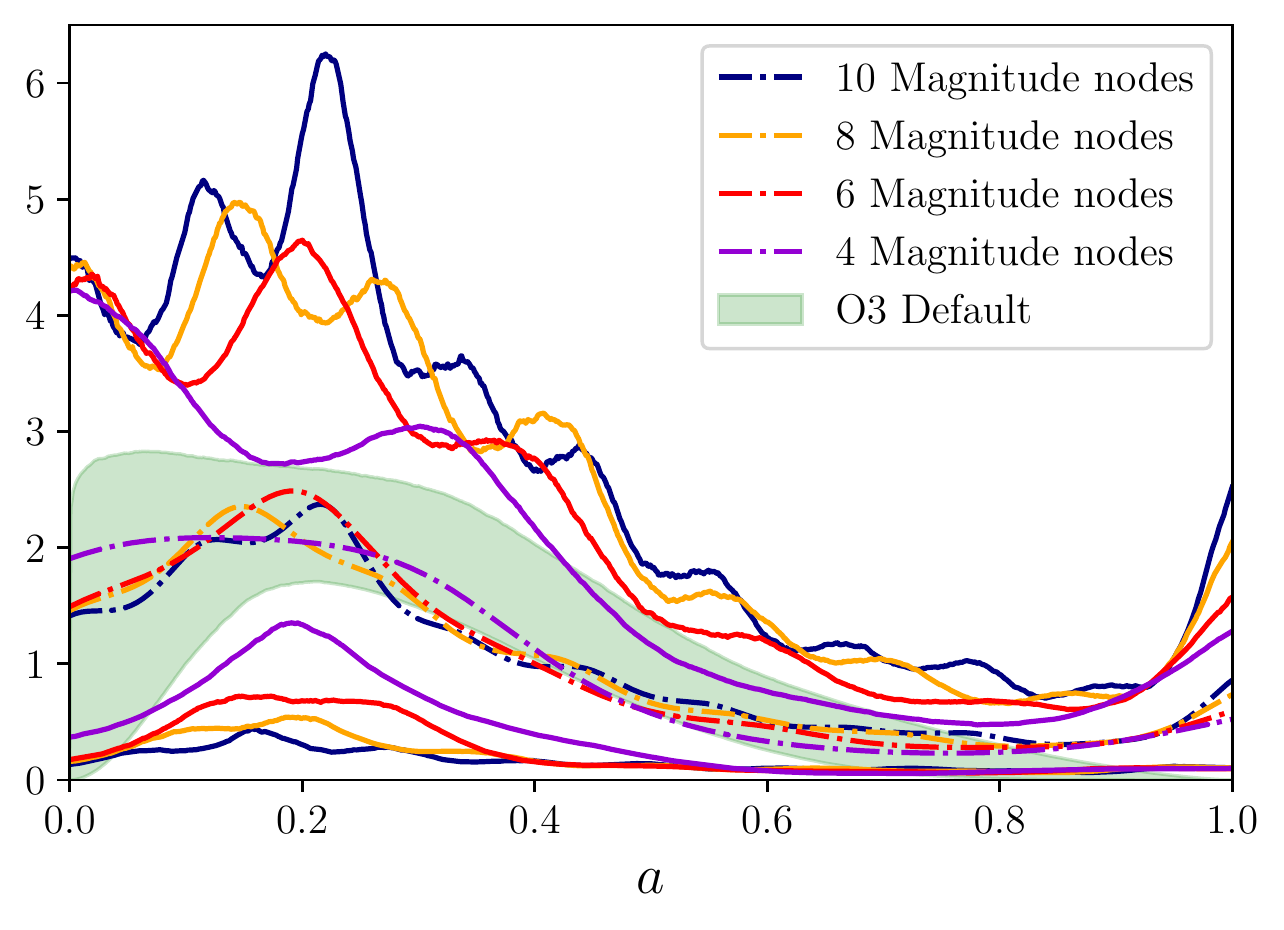}
  \caption{Distribution of spin magnitudes, with different numbers of nodes corresponding to different colors. All use 4 nodes in the tilt distribution. }
  \label{fig:spinmagnitudes4}
\end{figure}

\begin{figure}
  \includegraphics[width=\linewidth]{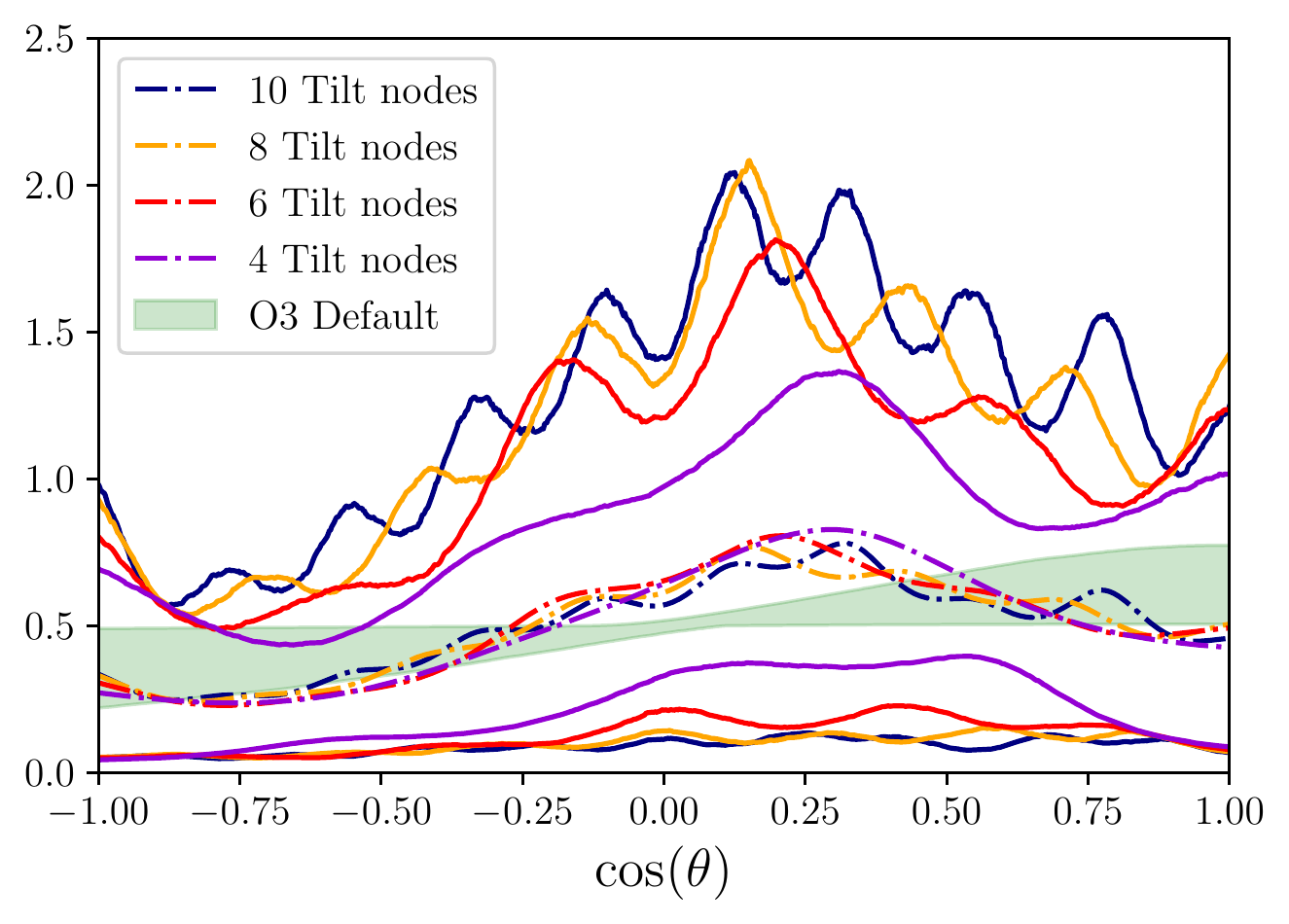}
  \caption{Distribution of spin tilts, with different numbers of nodes corresponding to different colors. All use 4 nodes in the spin magnitude distribution. }
  \label{fig:spintilts4}
\end{figure}

\section{Effect of priors on inferred distribution}\label{appendix:prior}
\begin{figure}
    \centering
    \includegraphics[width=\linewidth]{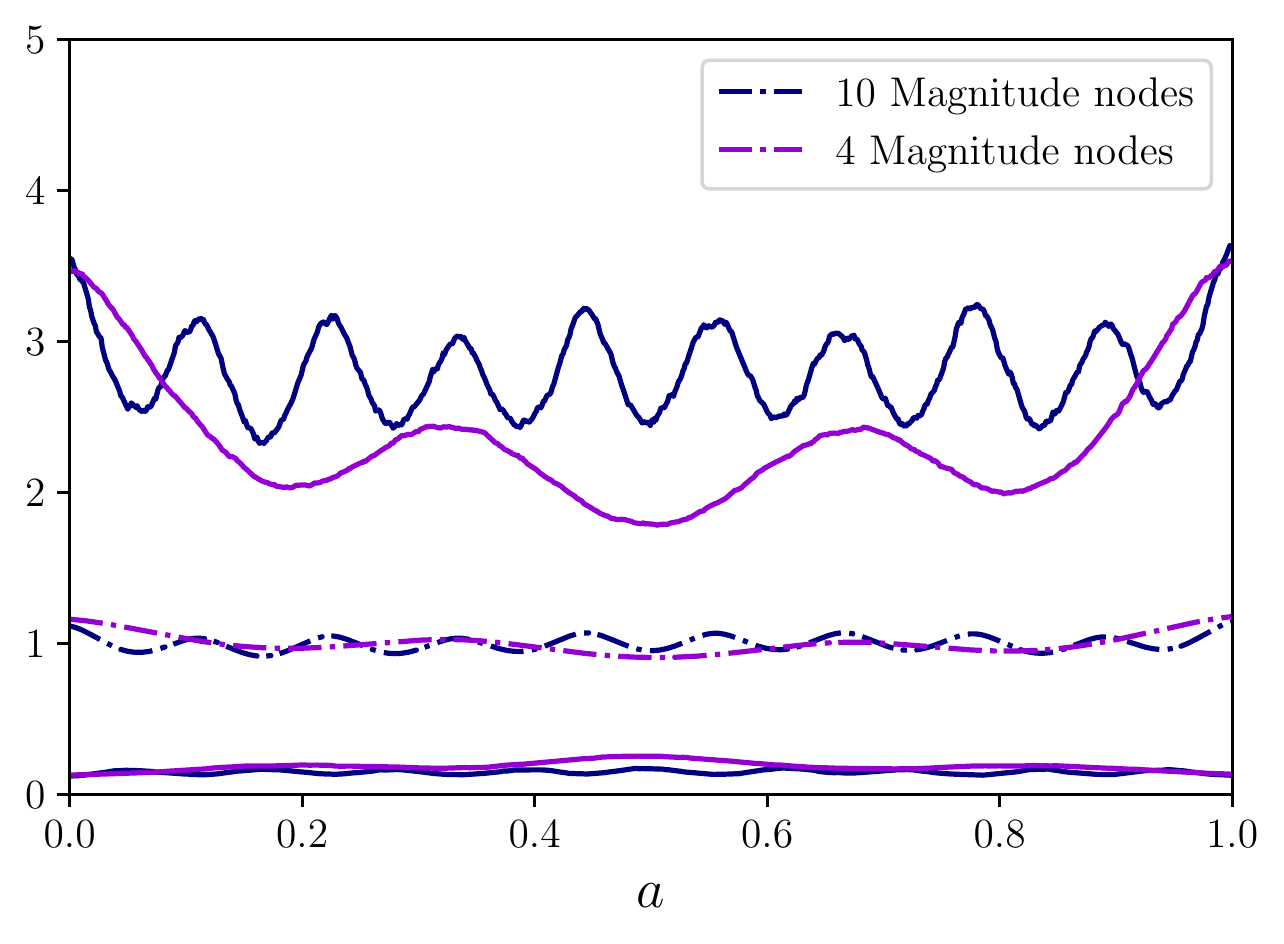}
    \includegraphics[width=\linewidth]{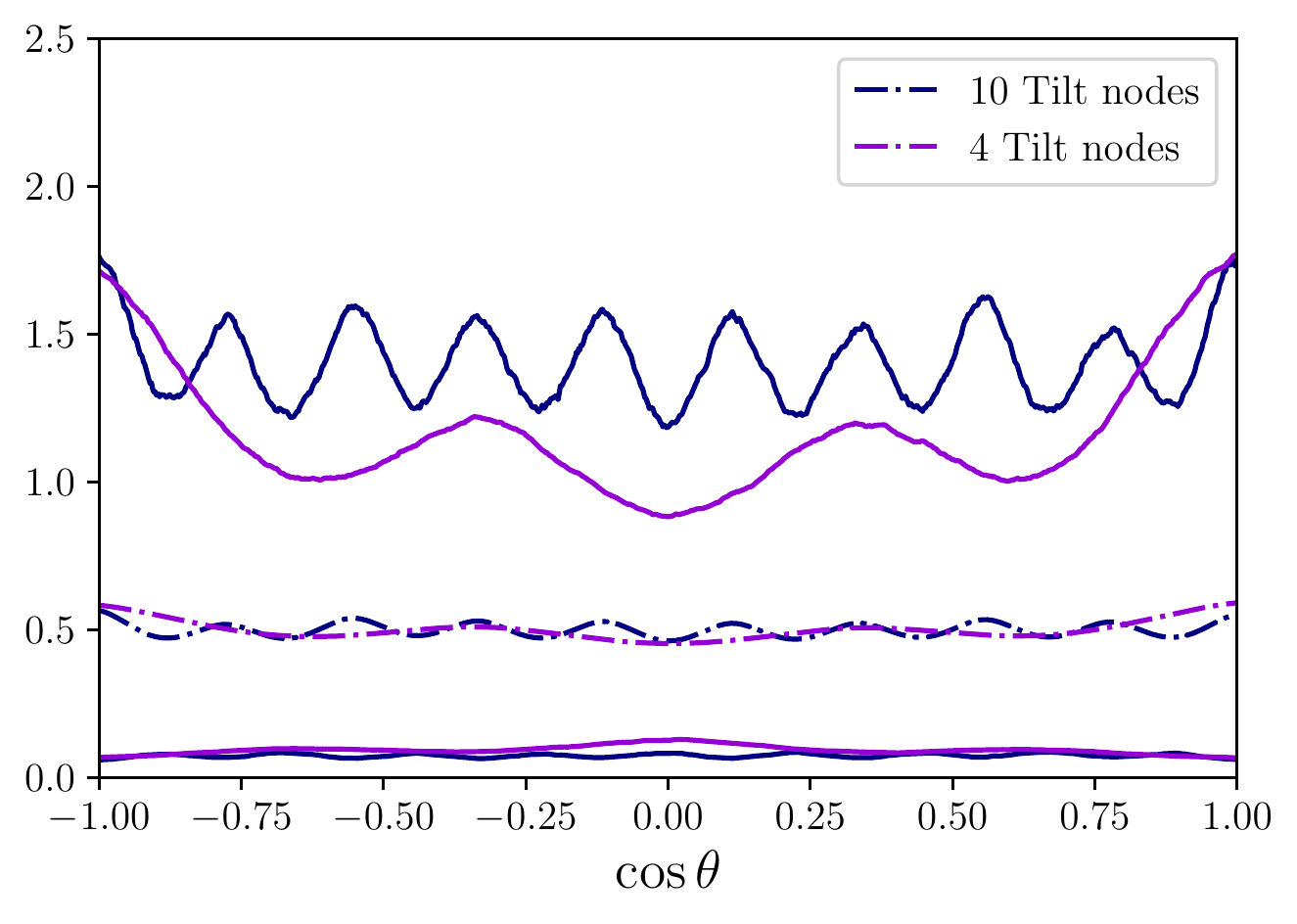}
    \caption{Comparison of the priors on the spin magnitude (top) and spin tilt (bottom) distributions. The envelopes (increased width in the distribution) are coincident with the node placement.}
    \label{fig:priorsonly}
\end{figure}
In this work, we adopt a prior on the spline node amplitudes that is a standard normal distribution. In this appendix, we show the results obtained using different choices of prior on the node amplitudes and different positions.

In Figure \ref{fig:priorsonly}, we show the distribution of spin magnitudes and tilts from prior draws only. We note that the average of the distribution is flat, reflecting the lack of any further structure imposed by the prior on the mean of the distribution. On the other hand, the upper limit of the 90\% credible regions shows considerable oscillations. These are coincident with the node locations and thus correspond to where the distribution is informed directly by the spline amplitude sample. These oscillations are thus an expected feature of the spline model. The regions in between these oscillations correspond to where the spline provides an interpolation between node locations.
 
\begin{figure}
    \centering
    \includegraphics[width=\linewidth]{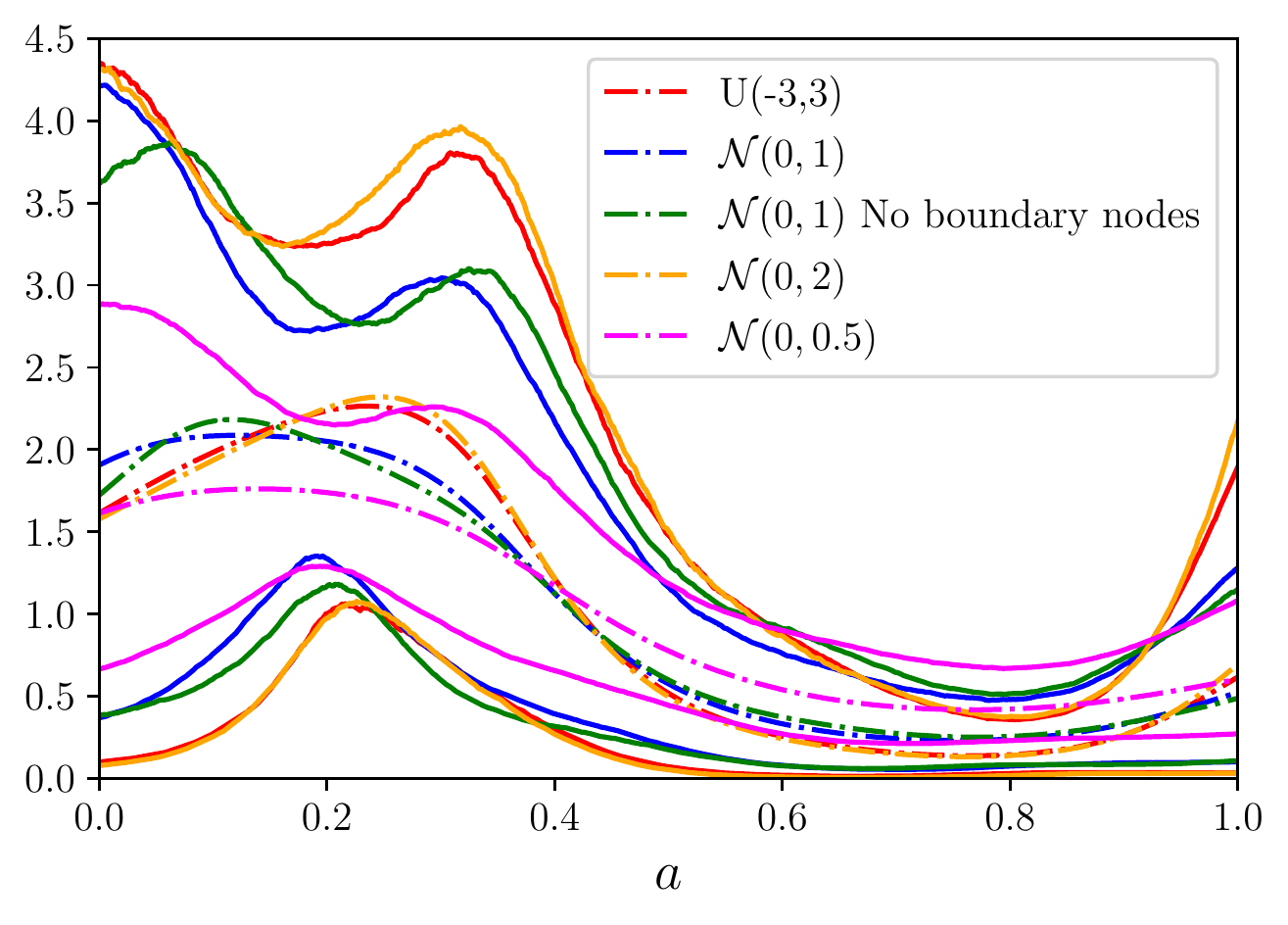}
    \includegraphics[width=\linewidth]{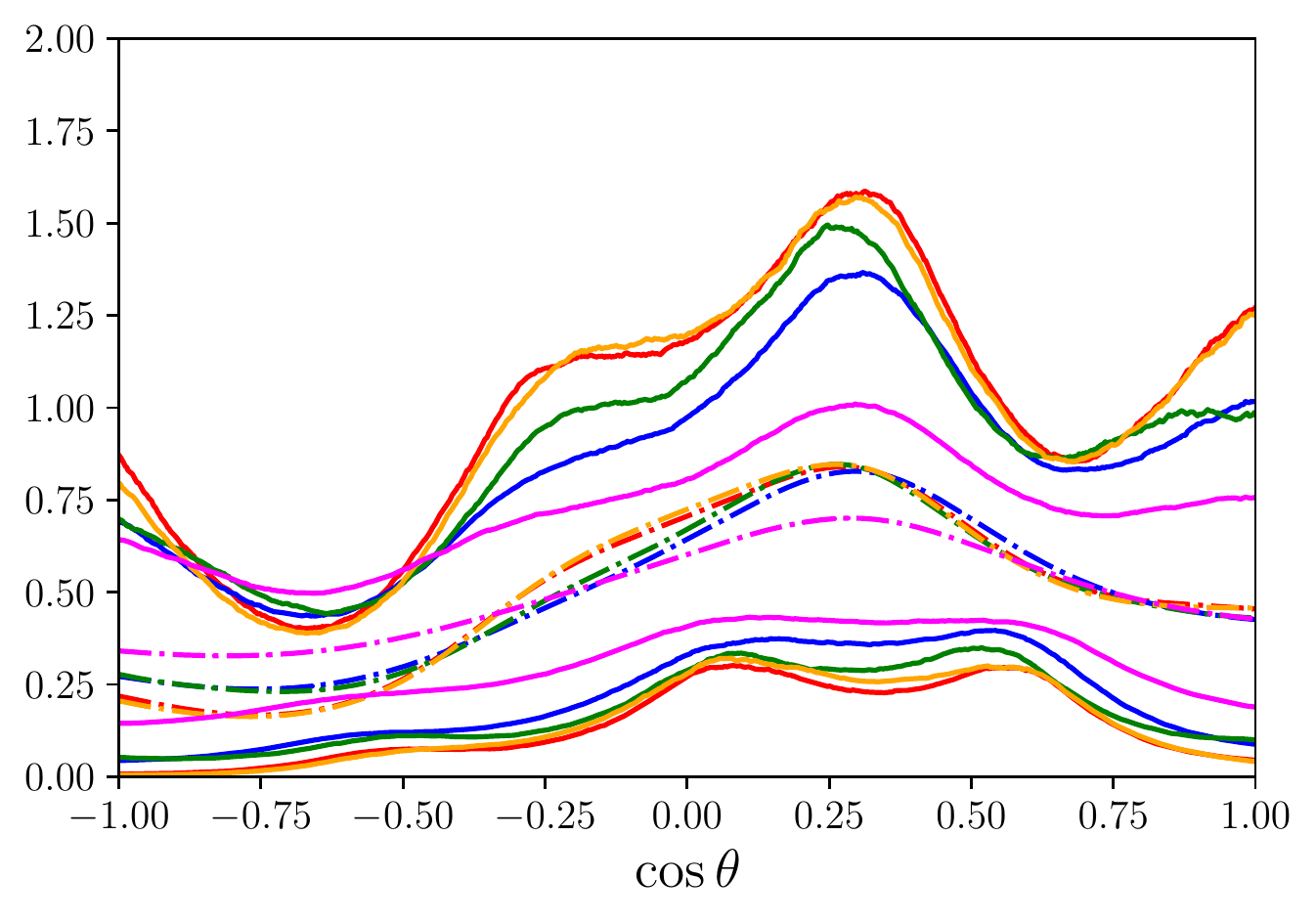}
    \caption{Comparison of spin reconstructions for different choices of priors on the spline nodes. The blue distribution corresponds to choice used in this work. The distribution ``No Boundary Nodes'' also places the boundaries linearly in the domain, but does not place any nodes on the domain boundary.}
    \label{fig:comparepriors}
\end{figure}

Figure \ref{fig:comparepriors} shows the inferred distribution for the 4 tilt node and 4 magnitude node model for three different choices of prior on the node amplitude and placement:
a broader Gaussian (yellow), a narrower Gaussian (magenta), a uniform distribution in $[-3, 3]$ and the unit Gaussian without the additional end nodes (see, Sec.~\ref{sec:splinepriors}).
They each result in comparable distributions within the statistical uncertainties, with the $\mathcal{N}(0, 0.5)$ prior giving the tightest constraints. 

\bibliography{body.bbl}

\begin{thebibliography}{66}%
\makeatletter
\providecommand \@ifxundefined [1]{%
 \@ifx{#1\undefined}
}%
\providecommand \@ifnum [1]{%
 \ifnum #1\expandafter \@firstoftwo
 \else \expandafter \@secondoftwo
 \fi
}%
\providecommand \@ifx [1]{%
 \ifx #1\expandafter \@firstoftwo
 \else \expandafter \@secondoftwo
 \fi
}%
\providecommand \natexlab [1]{#1}%
\providecommand \enquote  [1]{``#1''}%
\providecommand \bibnamefont  [1]{#1}%
\providecommand \bibfnamefont [1]{#1}%
\providecommand \citenamefont [1]{#1}%
\providecommand \href@noop [0]{\@secondoftwo}%
\providecommand \href [0]{\begingroup \@sanitize@url \@href}%
\providecommand \@href[1]{\@@startlink{#1}\@@href}%
\providecommand \@@href[1]{\endgroup#1\@@endlink}%
\providecommand \@sanitize@url [0]{\catcode `\\12\catcode `\$12\catcode
  `\&12\catcode `\#12\catcode `\^12\catcode `\_12\catcode `\%12\relax}%
\providecommand \@@startlink[1]{}%
\providecommand \@@endlink[0]{}%
\providecommand \url  [0]{\begingroup\@sanitize@url \@url }%
\providecommand \@url [1]{\endgroup\@href {#1}{\urlprefix }}%
\providecommand \urlprefix  [0]{URL }%
\providecommand \Eprint [0]{\href }%
\providecommand \doibase [0]{https://doi.org/}%
\providecommand \selectlanguage [0]{\@gobble}%
\providecommand \bibinfo  [0]{\@secondoftwo}%
\providecommand \bibfield  [0]{\@secondoftwo}%
\providecommand \translation [1]{[#1]}%
\providecommand \BibitemOpen [0]{}%
\providecommand \bibitemStop [0]{}%
\providecommand \bibitemNoStop [0]{.\EOS\space}%
\providecommand \EOS [0]{\spacefactor3000\relax}%
\providecommand \BibitemShut  [1]{\csname bibitem#1\endcsname}%
\let\auto@bib@innerbib\@empty
\bibitem [{\citenamefont {Abbott}\ \emph {et~al.}(2018)\citenamefont {Abbott}
  \emph {et~al.}}]{LVKdetctors}%
  \BibitemOpen
  \bibfield  {author} {\bibinfo {author} {\bibfnamefont {B.~P.}\ \bibnamefont
  {Abbott}} \emph {et~al.} (\bibinfo {collaboration} {KAGRA, LIGO Scientific,
  Virgo, VIRGO}),\ }\bibfield  {title} {\bibinfo {title} {{Prospects for
  observing and localizing gravitational-wave transients with Advanced LIGO,
  Advanced Virgo and KAGRA}},\ }\href
  {https://doi.org/10.1007/s41114-020-00026-9} {\bibfield  {journal} {\bibinfo
  {journal} {Living Rev. Rel.}\ }\textbf {\bibinfo {volume} {21}},\ \bibinfo
  {pages} {3} (\bibinfo {year} {2018})},\ \Eprint
  {https://arxiv.org/abs/1304.0670} {arXiv:1304.0670 [gr-qc]} \BibitemShut
  {NoStop}%
\bibitem [{\citenamefont {Aasi}\ \emph {et~al.}(2015)\citenamefont {Aasi} \emph
  {et~al.}}]{LIGO}%
  \BibitemOpen
  \bibfield  {author} {\bibinfo {author} {\bibfnamefont {J.}~\bibnamefont
  {Aasi}} \emph {et~al.} (\bibinfo {collaboration} {LIGO Scientific}),\
  }\bibfield  {title} {\bibinfo {title} {{Advanced LIGO}},\ }\href
  {https://doi.org/10.1088/0264-9381/32/7/074001} {\bibfield  {journal}
  {\bibinfo  {journal} {Class. Quant. Grav.}\ }\textbf {\bibinfo {volume}
  {32}},\ \bibinfo {pages} {074001} (\bibinfo {year} {2015})},\ \Eprint
  {https://arxiv.org/abs/1411.4547} {arXiv:1411.4547 [gr-qc]} \BibitemShut
  {NoStop}%
\bibitem [{\citenamefont {Acernese}\ \emph {et~al.}(2015)\citenamefont
  {Acernese} \emph {et~al.}}]{Virgo}%
  \BibitemOpen
  \bibfield  {author} {\bibinfo {author} {\bibfnamefont {F.}~\bibnamefont
  {Acernese}} \emph {et~al.} (\bibinfo {collaboration} {VIRGO}),\ }\bibfield
  {title} {\bibinfo {title} {{Advanced Virgo: a second-generation
  interferometric gravitational wave detector}},\ }\href
  {https://doi.org/10.1088/0264-9381/32/2/024001} {\bibfield  {journal}
  {\bibinfo  {journal} {Class. Quant. Grav.}\ }\textbf {\bibinfo {volume}
  {32}},\ \bibinfo {pages} {024001} (\bibinfo {year} {2015})},\ \Eprint
  {https://arxiv.org/abs/1408.3978} {arXiv:1408.3978 [gr-qc]} \BibitemShut
  {NoStop}%
\bibitem [{\citenamefont {{The LIGO Scientific Collaboration}}\ \emph
  {et~al.}(2021{\natexlab{a}})\citenamefont {{The LIGO Scientific
  Collaboration}}, \citenamefont {{the Virgo Collaboration}}, \citenamefont
  {{the KAGRA Collaboration}} \emph {et~al.}}]{gwtc3}%
  \BibitemOpen
  \bibfield  {author} {\bibinfo {author} {\bibnamefont {{The LIGO Scientific
  Collaboration}}}, \bibinfo {author} {\bibnamefont {{the Virgo
  Collaboration}}}, \bibinfo {author} {\bibnamefont {{the KAGRA
  Collaboration}}}, \emph {et~al.},\ }\bibfield  {title} {\bibinfo {title}
  {{GWTC-3: Compact Binary Coalescences Observed by LIGO and Virgo During the
  Second Part of the Third Observing Run}},\ }\href@noop {} {\bibfield
  {journal} {\bibinfo  {journal} {arXiv e-prints}\ ,\ \bibinfo {eid}
  {arXiv:2111.03606}} (\bibinfo {year} {2021}{\natexlab{a}})},\ \Eprint
  {https://arxiv.org/abs/2111.03606} {arXiv:2111.03606 [gr-qc]} \BibitemShut
  {NoStop}%
\bibitem [{\citenamefont {{LIGO Scientific Collaboration}}\ \emph
  {et~al.}(2019)\citenamefont {{LIGO Scientific Collaboration}}, \citenamefont
  {{Virgo Collaboration}} \emph {et~al.}}]{gwtc1}%
  \BibitemOpen
  \bibfield  {author} {\bibinfo {author} {\bibnamefont {{LIGO Scientific
  Collaboration}}}, \bibinfo {author} {\bibnamefont {{Virgo Collaboration}}},
  \emph {et~al.},\ }\bibfield  {title} {\bibinfo {title} {{GWTC-1: A
  Gravitational-Wave Transient Catalog of Compact Binary Mergers Observed by
  LIGO and Virgo during the First and Second Observing Runs}},\ }\href
  {https://doi.org/10.1103/PhysRevX.9.031040} {\bibfield  {journal} {\bibinfo
  {journal} {Physical Review X}\ }\textbf {\bibinfo {volume} {9}},\ \bibinfo
  {eid} {031040} (\bibinfo {year} {2019})},\ \Eprint
  {https://arxiv.org/abs/1811.12907} {arXiv:1811.12907 [astro-ph.HE]}
  \BibitemShut {NoStop}%
\bibitem [{\citenamefont {{The LIGO Scientific Collaboration}}\ \emph
  {et~al.}(2021{\natexlab{b}})\citenamefont {{The LIGO Scientific
  Collaboration}}, \citenamefont {{the Virgo Collaboration}}, \citenamefont
  {{the KAGRA Collaboration}} \emph {et~al.}}]{gwtc3pops}%
  \BibitemOpen
  \bibfield  {author} {\bibinfo {author} {\bibnamefont {{The LIGO Scientific
  Collaboration}}}, \bibinfo {author} {\bibnamefont {{the Virgo
  Collaboration}}}, \bibinfo {author} {\bibnamefont {{the KAGRA
  Collaboration}}}, \emph {et~al.},\ }\bibfield  {title} {\bibinfo {title}
  {{The population of merging compact binaries inferred using gravitational
  waves through GWTC-3}},\ }\href@noop {} {\bibfield  {journal} {\bibinfo
  {journal} {arXiv e-prints}\ ,\ \bibinfo {eid} {arXiv:2111.03634}} (\bibinfo
  {year} {2021}{\natexlab{b}})},\ \Eprint {https://arxiv.org/abs/2111.03634}
  {arXiv:2111.03634 [astro-ph.HE]} \BibitemShut {NoStop}%
\bibitem [{\citenamefont {{LIGO Scientific Collaboration}}\ \emph
  {et~al.}(2021{\natexlab{a}})\citenamefont {{LIGO Scientific Collaboration}},
  \citenamefont {{Virgo Collaboration}} \emph {et~al.}}]{gwtc2pops}%
  \BibitemOpen
  \bibfield  {author} {\bibinfo {author} {\bibnamefont {{LIGO Scientific
  Collaboration}}}, \bibinfo {author} {\bibnamefont {{Virgo Collaboration}}},
  \emph {et~al.},\ }\bibfield  {title} {\bibinfo {title} {{Population
  Properties of Compact Objects from the Second LIGO-Virgo Gravitational-Wave
  Transient Catalog}},\ }\href {https://doi.org/10.3847/2041-8213/abe949}
  {\bibfield  {journal} {\bibinfo  {journal} {\apjl}\ }\textbf {\bibinfo
  {volume} {913}},\ \bibinfo {eid} {L7} (\bibinfo {year}
  {2021}{\natexlab{a}})},\ \Eprint {https://arxiv.org/abs/2010.14533}
  {arXiv:2010.14533 [astro-ph.HE]} \BibitemShut {NoStop}%
\bibitem [{\citenamefont {{The LIGO Scientific Collaboration}}\ \emph
  {et~al.}(2021{\natexlab{c}})\citenamefont {{The LIGO Scientific
  Collaboration}}, \citenamefont {{the Virgo Collaboration}}, \citenamefont
  {{the KAGRA Collaboration}} \emph {et~al.}}]{GWTC3Cosmology}%
  \BibitemOpen
  \bibfield  {author} {\bibinfo {author} {\bibnamefont {{The LIGO Scientific
  Collaboration}}}, \bibinfo {author} {\bibnamefont {{the Virgo
  Collaboration}}}, \bibinfo {author} {\bibnamefont {{the KAGRA
  Collaboration}}}, \emph {et~al.},\ }\bibfield  {title} {\bibinfo {title}
  {{Constraints on the cosmic expansion history from GWTC-3}},\ }\href@noop {}
  {\bibfield  {journal} {\bibinfo  {journal} {arXiv e-prints}\ ,\ \bibinfo
  {eid} {arXiv:2111.03604}} (\bibinfo {year} {2021}{\natexlab{c}})},\ \Eprint
  {https://arxiv.org/abs/2111.03604} {arXiv:2111.03604 [astro-ph.CO]}
  \BibitemShut {NoStop}%
\bibitem [{\citenamefont {{The LIGO Scientific Collaboration}}\ \emph
  {et~al.}(2021{\natexlab{d}})\citenamefont {{The LIGO Scientific
  Collaboration}}, \citenamefont {{the Virgo Collaboration}}, \citenamefont
  {{the KAGRA Collaboration}} \emph {et~al.}}]{GWTC3TGR}%
  \BibitemOpen
  \bibfield  {author} {\bibinfo {author} {\bibnamefont {{The LIGO Scientific
  Collaboration}}}, \bibinfo {author} {\bibnamefont {{the Virgo
  Collaboration}}}, \bibinfo {author} {\bibnamefont {{the KAGRA
  Collaboration}}}, \emph {et~al.},\ }\bibfield  {title} {\bibinfo {title}
  {{Tests of General Relativity with GWTC-3}},\ }\href@noop {} {\bibfield
  {journal} {\bibinfo  {journal} {arXiv e-prints}\ ,\ \bibinfo {eid}
  {arXiv:2112.06861}} (\bibinfo {year} {2021}{\natexlab{d}})},\ \Eprint
  {https://arxiv.org/abs/2112.06861} {arXiv:2112.06861 [gr-qc]} \BibitemShut
  {NoStop}%
\bibitem [{\citenamefont {{Kimball}}\ \emph {et~al.}(2021)\citenamefont
  {{Kimball}}, \citenamefont {{Talbot}}, \citenamefont {{Berry}}, \citenamefont
  {{Zevin}}, \citenamefont {{Thrane}}, \citenamefont {{Kalogera}},
  \citenamefont {{Buscicchio}}, \citenamefont {{Carney}}, \citenamefont
  {{Dent}}, \citenamefont {{Middleton}}, \citenamefont {{Payne}}, \citenamefont
  {{Veitch}},\ and\ \citenamefont {{Williams}}}]{Kimball21}%
  \BibitemOpen
  \bibfield  {author} {\bibinfo {author} {\bibfnamefont {C.}~\bibnamefont
  {{Kimball}}}, \bibinfo {author} {\bibfnamefont {C.}~\bibnamefont {{Talbot}}},
  \bibinfo {author} {\bibfnamefont {C.~P.~L.}\ \bibnamefont {{Berry}}},
  \bibinfo {author} {\bibfnamefont {M.}~\bibnamefont {{Zevin}}}, \bibinfo
  {author} {\bibfnamefont {E.}~\bibnamefont {{Thrane}}}, \bibinfo {author}
  {\bibfnamefont {V.}~\bibnamefont {{Kalogera}}}, \bibinfo {author}
  {\bibfnamefont {R.}~\bibnamefont {{Buscicchio}}}, \bibinfo {author}
  {\bibfnamefont {M.}~\bibnamefont {{Carney}}}, \bibinfo {author}
  {\bibfnamefont {T.}~\bibnamefont {{Dent}}}, \bibinfo {author} {\bibfnamefont
  {H.}~\bibnamefont {{Middleton}}}, \bibinfo {author} {\bibfnamefont
  {E.}~\bibnamefont {{Payne}}}, \bibinfo {author} {\bibfnamefont
  {J.}~\bibnamefont {{Veitch}}},\ and\ \bibinfo {author} {\bibfnamefont
  {D.}~\bibnamefont {{Williams}}},\ }\bibfield  {title} {\bibinfo {title}
  {{Evidence for Hierarchical Black Hole Mergers in the Second LIGO-Virgo
  Gravitational Wave Catalog}},\ }\href
  {https://doi.org/10.3847/2041-8213/ac0aef} {\bibfield  {journal} {\bibinfo
  {journal} {\apjl}\ }\textbf {\bibinfo {volume} {915}},\ \bibinfo {eid} {L35}
  (\bibinfo {year} {2021})},\ \Eprint {https://arxiv.org/abs/2011.05332}
  {arXiv:2011.05332 [astro-ph.HE]} \BibitemShut {NoStop}%
\bibitem [{\citenamefont {{Fishbach}}\ \emph {et~al.}(2022)\citenamefont
  {{Fishbach}}, \citenamefont {{Kimball}},\ and\ \citenamefont
  {{Kalogera}}}]{Fishbach22}%
  \BibitemOpen
  \bibfield  {author} {\bibinfo {author} {\bibfnamefont {M.}~\bibnamefont
  {{Fishbach}}}, \bibinfo {author} {\bibfnamefont {C.}~\bibnamefont
  {{Kimball}}},\ and\ \bibinfo {author} {\bibfnamefont {V.}~\bibnamefont
  {{Kalogera}}},\ }\bibfield  {title} {\bibinfo {title} {{Limits on
  Hierarchical Black Hole Mergers from the Most Negative {\ensuremath{\chi}}
  $_{eff}$ Systems}},\ }\href {https://doi.org/10.3847/2041-8213/ac86c4}
  {\bibfield  {journal} {\bibinfo  {journal} {\apjl}\ }\textbf {\bibinfo
  {volume} {935}},\ \bibinfo {eid} {L26} (\bibinfo {year} {2022})},\ \Eprint
  {https://arxiv.org/abs/2207.02924} {arXiv:2207.02924 [astro-ph.HE]}
  \BibitemShut {NoStop}%
\bibitem [{\citenamefont {{Gerosa}}\ and\ \citenamefont
  {{Fishbach}}(2021)}]{GerosaFishbach}%
  \BibitemOpen
  \bibfield  {author} {\bibinfo {author} {\bibfnamefont {D.}~\bibnamefont
  {{Gerosa}}}\ and\ \bibinfo {author} {\bibfnamefont {M.}~\bibnamefont
  {{Fishbach}}},\ }\bibfield  {title} {\bibinfo {title} {{Hierarchical mergers
  of stellar-mass black holes and their gravitational-wave signatures}},\
  }\href {https://doi.org/10.1038/s41550-021-01398-w} {\bibfield  {journal}
  {\bibinfo  {journal} {Nature Astronomy}\ }\textbf {\bibinfo {volume} {5}},\
  \bibinfo {pages} {749} (\bibinfo {year} {2021})},\ \Eprint
  {https://arxiv.org/abs/2105.03439} {arXiv:2105.03439 [astro-ph.HE]}
  \BibitemShut {NoStop}%
\bibitem [{\citenamefont {{Fishbach}}\ \emph {et~al.}(2017)\citenamefont
  {{Fishbach}}, \citenamefont {{Holz}},\ and\ \citenamefont
  {{Farr}}}]{Fishbach17}%
  \BibitemOpen
  \bibfield  {author} {\bibinfo {author} {\bibfnamefont {M.}~\bibnamefont
  {{Fishbach}}}, \bibinfo {author} {\bibfnamefont {D.~E.}\ \bibnamefont
  {{Holz}}},\ and\ \bibinfo {author} {\bibfnamefont {B.}~\bibnamefont
  {{Farr}}},\ }\bibfield  {title} {\bibinfo {title} {{Are LIGO's Black Holes
  Made from Smaller Black Holes?}},\ }\href
  {https://doi.org/10.3847/2041-8213/aa7045} {\bibfield  {journal} {\bibinfo
  {journal} {\apjl}\ }\textbf {\bibinfo {volume} {840}},\ \bibinfo {eid} {L24}
  (\bibinfo {year} {2017})},\ \Eprint {https://arxiv.org/abs/1703.06869}
  {arXiv:1703.06869 [astro-ph.HE]} \BibitemShut {NoStop}%
\bibitem [{\citenamefont {{Fuller}}\ and\ \citenamefont
  {{Ma}}(2019)}]{FullerMa15}%
  \BibitemOpen
  \bibfield  {author} {\bibinfo {author} {\bibfnamefont {J.}~\bibnamefont
  {{Fuller}}}\ and\ \bibinfo {author} {\bibfnamefont {L.}~\bibnamefont
  {{Ma}}},\ }\bibfield  {title} {\bibinfo {title} {{Most Black Holes Are Born
  Very Slowly Rotating}},\ }\href {https://doi.org/10.3847/2041-8213/ab339b}
  {\bibfield  {journal} {\bibinfo  {journal} {\apjl}\ }\textbf {\bibinfo
  {volume} {881}},\ \bibinfo {eid} {L1} (\bibinfo {year} {2019})},\ \Eprint
  {https://arxiv.org/abs/1907.03714} {arXiv:1907.03714 [astro-ph.SR]}
  \BibitemShut {NoStop}%
\bibitem [{\citenamefont {{Qin}}\ \emph {et~al.}(2019)\citenamefont {{Qin}},
  \citenamefont {{Marchant}}, \citenamefont {{Fragos}}, \citenamefont
  {{Meynet}},\ and\ \citenamefont {{Kalogera}}}]{Qin}%
  \BibitemOpen
  \bibfield  {author} {\bibinfo {author} {\bibfnamefont {Y.}~\bibnamefont
  {{Qin}}}, \bibinfo {author} {\bibfnamefont {P.}~\bibnamefont {{Marchant}}},
  \bibinfo {author} {\bibfnamefont {T.}~\bibnamefont {{Fragos}}}, \bibinfo
  {author} {\bibfnamefont {G.}~\bibnamefont {{Meynet}}},\ and\ \bibinfo
  {author} {\bibfnamefont {V.}~\bibnamefont {{Kalogera}}},\ }\bibfield  {title}
  {\bibinfo {title} {{On the Origin of Black Hole Spin in High-mass X-Ray
  Binaries}},\ }\href {https://doi.org/10.3847/2041-8213/aaf97b} {\bibfield
  {journal} {\bibinfo  {journal} {\apjl}\ }\textbf {\bibinfo {volume} {870}},\
  \bibinfo {eid} {L18} (\bibinfo {year} {2019})},\ \Eprint
  {https://arxiv.org/abs/1810.13016} {arXiv:1810.13016 [astro-ph.SR]}
  \BibitemShut {NoStop}%
\bibitem [{\citenamefont {{Heger}}\ \emph {et~al.}(2005)\citenamefont
  {{Heger}}, \citenamefont {{Woosley}},\ and\ \citenamefont
  {{Spruit}}}]{Heger}%
  \BibitemOpen
  \bibfield  {author} {\bibinfo {author} {\bibfnamefont {A.}~\bibnamefont
  {{Heger}}}, \bibinfo {author} {\bibfnamefont {S.~E.}\ \bibnamefont
  {{Woosley}}},\ and\ \bibinfo {author} {\bibfnamefont {H.~C.}\ \bibnamefont
  {{Spruit}}},\ }\bibfield  {title} {\bibinfo {title} {{Presupernova Evolution
  of Differentially Rotating Massive Stars Including Magnetic Fields}},\ }\href
  {https://doi.org/10.1086/429868} {\bibfield  {journal} {\bibinfo  {journal}
  {\apj}\ }\textbf {\bibinfo {volume} {626}},\ \bibinfo {pages} {350} (\bibinfo
  {year} {2005})},\ \Eprint {https://arxiv.org/abs/astro-ph/0409422}
  {arXiv:astro-ph/0409422 [astro-ph]} \BibitemShut {NoStop}%
\bibitem [{\citenamefont {{Zevin}}\ and\ \citenamefont
  {{Bavera}}(2022)}]{Zevin22}%
  \BibitemOpen
  \bibfield  {author} {\bibinfo {author} {\bibfnamefont {M.}~\bibnamefont
  {{Zevin}}}\ and\ \bibinfo {author} {\bibfnamefont {S.~S.}\ \bibnamefont
  {{Bavera}}},\ }\bibfield  {title} {\bibinfo {title} {{Suspicious Siblings:
  The Distribution of Mass and Spin across Component Black Holes in Isolated
  Binary Evolution}},\ }\href {https://doi.org/10.3847/1538-4357/ac6f5d}
  {\bibfield  {journal} {\bibinfo  {journal} {\apj}\ }\textbf {\bibinfo
  {volume} {933}},\ \bibinfo {eid} {86} (\bibinfo {year} {2022})},\ \Eprint
  {https://arxiv.org/abs/2203.02515} {arXiv:2203.02515 [astro-ph.HE]}
  \BibitemShut {NoStop}%
\bibitem [{\citenamefont {{Olejak}}\ and\ \citenamefont
  {{Belczynski}}(2021)}]{Olejak}%
  \BibitemOpen
  \bibfield  {author} {\bibinfo {author} {\bibfnamefont {A.}~\bibnamefont
  {{Olejak}}}\ and\ \bibinfo {author} {\bibfnamefont {K.}~\bibnamefont
  {{Belczynski}}},\ }\bibfield  {title} {\bibinfo {title} {{The Implications of
  High Black Hole Spins for the Origin of Binary Black Hole Mergers}},\ }\href
  {https://doi.org/10.3847/2041-8213/ac2f48} {\bibfield  {journal} {\bibinfo
  {journal} {\apjl}\ }\textbf {\bibinfo {volume} {921}},\ \bibinfo {eid} {L2}
  (\bibinfo {year} {2021})},\ \Eprint {https://arxiv.org/abs/2109.06872}
  {arXiv:2109.06872 [astro-ph.HE]} \BibitemShut {NoStop}%
\bibitem [{\citenamefont {{Bavera}}\ \emph {et~al.}(2020)\citenamefont
  {{Bavera}}, \citenamefont {{Fragos}}, \citenamefont {{Qin}}, \citenamefont
  {{Zapartas}}, \citenamefont {{Neijssel}}, \citenamefont {{Mandel}},
  \citenamefont {{Batta}}, \citenamefont {{Gaebel}}, \citenamefont
  {{Kimball}},\ and\ \citenamefont {{Stevenson}}}]{Bavera20}%
  \BibitemOpen
  \bibfield  {author} {\bibinfo {author} {\bibfnamefont {S.~S.}\ \bibnamefont
  {{Bavera}}}, \bibinfo {author} {\bibfnamefont {T.}~\bibnamefont {{Fragos}}},
  \bibinfo {author} {\bibfnamefont {Y.}~\bibnamefont {{Qin}}}, \bibinfo
  {author} {\bibfnamefont {E.}~\bibnamefont {{Zapartas}}}, \bibinfo {author}
  {\bibfnamefont {C.~J.}\ \bibnamefont {{Neijssel}}}, \bibinfo {author}
  {\bibfnamefont {I.}~\bibnamefont {{Mandel}}}, \bibinfo {author}
  {\bibfnamefont {A.}~\bibnamefont {{Batta}}}, \bibinfo {author} {\bibfnamefont
  {S.~M.}\ \bibnamefont {{Gaebel}}}, \bibinfo {author} {\bibfnamefont
  {C.}~\bibnamefont {{Kimball}}},\ and\ \bibinfo {author} {\bibfnamefont
  {S.}~\bibnamefont {{Stevenson}}},\ }\bibfield  {title} {\bibinfo {title}
  {{The origin of spin in binary black holes. Predicting the distributions of
  the main observables of Advanced LIGO}},\ }\href
  {https://doi.org/10.1051/0004-6361/201936204} {\bibfield  {journal} {\bibinfo
   {journal} {\aap}\ }\textbf {\bibinfo {volume} {635}},\ \bibinfo {eid} {A97}
  (\bibinfo {year} {2020})},\ \Eprint {https://arxiv.org/abs/1906.12257}
  {arXiv:1906.12257 [astro-ph.HE]} \BibitemShut {NoStop}%
\bibitem [{\citenamefont {{Bavera}}\ \emph {et~al.}(2021)\citenamefont
  {{Bavera}}, \citenamefont {{Fragos}}, \citenamefont {{Zevin}}, \citenamefont
  {{Berry}}, \citenamefont {{Marchant}}, \citenamefont {{Andrews}},
  \citenamefont {{Coughlin}}, \citenamefont {{Dotter}}, \citenamefont
  {{Kovlakas}}, \citenamefont {{Misra}}, \citenamefont {{Serra-Perez}},
  \citenamefont {{Qin}}, \citenamefont {{Rocha}}, \citenamefont
  {{Rom{\'a}n-Garza}}, \citenamefont {{Tran}},\ and\ \citenamefont
  {{Zapartas}}}]{Bavera21}%
  \BibitemOpen
  \bibfield  {author} {\bibinfo {author} {\bibfnamefont {S.~S.}\ \bibnamefont
  {{Bavera}}}, \bibinfo {author} {\bibfnamefont {T.}~\bibnamefont {{Fragos}}},
  \bibinfo {author} {\bibfnamefont {M.}~\bibnamefont {{Zevin}}}, \bibinfo
  {author} {\bibfnamefont {C.~P.~L.}\ \bibnamefont {{Berry}}}, \bibinfo
  {author} {\bibfnamefont {P.}~\bibnamefont {{Marchant}}}, \bibinfo {author}
  {\bibfnamefont {J.~J.}\ \bibnamefont {{Andrews}}}, \bibinfo {author}
  {\bibfnamefont {S.}~\bibnamefont {{Coughlin}}}, \bibinfo {author}
  {\bibfnamefont {A.}~\bibnamefont {{Dotter}}}, \bibinfo {author}
  {\bibfnamefont {K.}~\bibnamefont {{Kovlakas}}}, \bibinfo {author}
  {\bibfnamefont {D.}~\bibnamefont {{Misra}}}, \bibinfo {author} {\bibfnamefont
  {J.~G.}\ \bibnamefont {{Serra-Perez}}}, \bibinfo {author} {\bibfnamefont
  {Y.}~\bibnamefont {{Qin}}}, \bibinfo {author} {\bibfnamefont {K.~A.}\
  \bibnamefont {{Rocha}}}, \bibinfo {author} {\bibfnamefont {J.}~\bibnamefont
  {{Rom{\'a}n-Garza}}}, \bibinfo {author} {\bibfnamefont {N.~H.}\ \bibnamefont
  {{Tran}}},\ and\ \bibinfo {author} {\bibfnamefont {E.}~\bibnamefont
  {{Zapartas}}},\ }\bibfield  {title} {\bibinfo {title} {{The impact of
  mass-transfer physics on the observable properties of field binary black hole
  populations}},\ }\href {https://doi.org/10.1051/0004-6361/202039804}
  {\bibfield  {journal} {\bibinfo  {journal} {\aap}\ }\textbf {\bibinfo
  {volume} {647}},\ \bibinfo {eid} {A153} (\bibinfo {year} {2021})},\ \Eprint
  {https://arxiv.org/abs/2010.16333} {arXiv:2010.16333 [astro-ph.HE]}
  \BibitemShut {NoStop}%
\bibitem [{\citenamefont {{Hinder}}\ \emph {et~al.}(2008)\citenamefont
  {{Hinder}}, \citenamefont {{Vaishnav}}, \citenamefont {{Herrmann}},
  \citenamefont {{Shoemaker}},\ and\ \citenamefont {{Laguna}}}]{Hinder}%
  \BibitemOpen
  \bibfield  {author} {\bibinfo {author} {\bibfnamefont {I.}~\bibnamefont
  {{Hinder}}}, \bibinfo {author} {\bibfnamefont {B.}~\bibnamefont
  {{Vaishnav}}}, \bibinfo {author} {\bibfnamefont {F.}~\bibnamefont
  {{Herrmann}}}, \bibinfo {author} {\bibfnamefont {D.~M.}\ \bibnamefont
  {{Shoemaker}}},\ and\ \bibinfo {author} {\bibfnamefont {P.}~\bibnamefont
  {{Laguna}}},\ }\bibfield  {title} {\bibinfo {title} {{Circularization and
  final spin in eccentric binary-black-hole inspirals}},\ }\href
  {https://doi.org/10.1103/PhysRevD.77.081502} {\bibfield  {journal} {\bibinfo
  {journal} {\prd}\ }\textbf {\bibinfo {volume} {77}},\ \bibinfo {eid} {081502}
  (\bibinfo {year} {2008})},\ \Eprint {https://arxiv.org/abs/0710.5167}
  {arXiv:0710.5167 [gr-qc]} \BibitemShut {NoStop}%
\bibitem [{\citenamefont {{Hofmann}}\ \emph {et~al.}(2016)\citenamefont
  {{Hofmann}}, \citenamefont {{Barausse}},\ and\ \citenamefont
  {{Rezzolla}}}]{Hofmann}%
  \BibitemOpen
  \bibfield  {author} {\bibinfo {author} {\bibfnamefont {F.}~\bibnamefont
  {{Hofmann}}}, \bibinfo {author} {\bibfnamefont {E.}~\bibnamefont
  {{Barausse}}},\ and\ \bibinfo {author} {\bibfnamefont {L.}~\bibnamefont
  {{Rezzolla}}},\ }\bibfield  {title} {\bibinfo {title} {{The Final Spin from
  Binary Black Holes in Quasi-circular Orbits}},\ }\href
  {https://doi.org/10.3847/2041-8205/825/2/L19} {\bibfield  {journal} {\bibinfo
   {journal} {\apjl}\ }\textbf {\bibinfo {volume} {825}},\ \bibinfo {eid} {L19}
  (\bibinfo {year} {2016})},\ \Eprint {https://arxiv.org/abs/1605.01938}
  {arXiv:1605.01938 [gr-qc]} \BibitemShut {NoStop}%
\bibitem [{\citenamefont {{Farr}}\ \emph {et~al.}(2017)\citenamefont {{Farr}},
  \citenamefont {{Stevenson}}, \citenamefont {{Miller}}, \citenamefont
  {{Mandel}}, \citenamefont {{Farr}},\ and\ \citenamefont
  {{Vecchio}}}]{Farr17}%
  \BibitemOpen
  \bibfield  {author} {\bibinfo {author} {\bibfnamefont {W.~M.}\ \bibnamefont
  {{Farr}}}, \bibinfo {author} {\bibfnamefont {S.}~\bibnamefont {{Stevenson}}},
  \bibinfo {author} {\bibfnamefont {M.~C.}\ \bibnamefont {{Miller}}}, \bibinfo
  {author} {\bibfnamefont {I.}~\bibnamefont {{Mandel}}}, \bibinfo {author}
  {\bibfnamefont {B.}~\bibnamefont {{Farr}}},\ and\ \bibinfo {author}
  {\bibfnamefont {A.}~\bibnamefont {{Vecchio}}},\ }\bibfield  {title} {\bibinfo
  {title} {{Distinguishing spin-aligned and isotropic black hole populations
  with gravitational waves}},\ }\href {https://doi.org/10.1038/nature23453}
  {\bibfield  {journal} {\bibinfo  {journal} {\nat}\ }\textbf {\bibinfo
  {volume} {548}},\ \bibinfo {pages} {426} (\bibinfo {year} {2017})},\ \Eprint
  {https://arxiv.org/abs/1706.01385} {arXiv:1706.01385 [astro-ph.HE]}
  \BibitemShut {NoStop}%
\bibitem [{\citenamefont {{Kalogera}}(2000)}]{Kalogera2000}%
  \BibitemOpen
  \bibfield  {author} {\bibinfo {author} {\bibfnamefont {V.}~\bibnamefont
  {{Kalogera}}},\ }\bibfield  {title} {\bibinfo {title} {{Spin-Orbit
  Misalignment in Close Binaries with Two Compact Objects}},\ }\href
  {https://doi.org/10.1086/309400} {\bibfield  {journal} {\bibinfo  {journal}
  {\apj}\ }\textbf {\bibinfo {volume} {541}},\ \bibinfo {pages} {319} (\bibinfo
  {year} {2000})},\ \Eprint {https://arxiv.org/abs/astro-ph/9911417}
  {arXiv:astro-ph/9911417 [astro-ph]} \BibitemShut {NoStop}%
\bibitem [{\citenamefont {{Mandel}}\ and\ \citenamefont
  {{O'Shaughnessy}}(2010)}]{mandel10}%
  \BibitemOpen
  \bibfield  {author} {\bibinfo {author} {\bibfnamefont {I.}~\bibnamefont
  {{Mandel}}}\ and\ \bibinfo {author} {\bibfnamefont {R.}~\bibnamefont
  {{O'Shaughnessy}}},\ }\bibfield  {title} {\bibinfo {title} {{Compact binary
  coalescences in the band of ground-based gravitational-wave detectors}},\
  }\href {https://doi.org/10.1088/0264-9381/27/11/114007} {\bibfield  {journal}
  {\bibinfo  {journal} {Classical and Quantum Gravity}\ }\textbf {\bibinfo
  {volume} {27}},\ \bibinfo {eid} {114007} (\bibinfo {year} {2010})},\ \Eprint
  {https://arxiv.org/abs/0912.1074} {arXiv:0912.1074 [astro-ph.HE]}
  \BibitemShut {NoStop}%
\bibitem [{\citenamefont {{Rodriguez}}\ \emph {et~al.}(2016)\citenamefont
  {{Rodriguez}}, \citenamefont {{Zevin}}, \citenamefont {{Pankow}},
  \citenamefont {{Kalogera}},\ and\ \citenamefont {{Rasio}}}]{Rodriguez}%
  \BibitemOpen
  \bibfield  {author} {\bibinfo {author} {\bibfnamefont {C.~L.}\ \bibnamefont
  {{Rodriguez}}}, \bibinfo {author} {\bibfnamefont {M.}~\bibnamefont
  {{Zevin}}}, \bibinfo {author} {\bibfnamefont {C.}~\bibnamefont {{Pankow}}},
  \bibinfo {author} {\bibfnamefont {V.}~\bibnamefont {{Kalogera}}},\ and\
  \bibinfo {author} {\bibfnamefont {F.~A.}\ \bibnamefont {{Rasio}}},\
  }\bibfield  {title} {\bibinfo {title} {{Illuminating Black Hole Binary
  Formation Channels with Spins in Advanced LIGO}},\ }\href
  {https://doi.org/10.3847/2041-8205/832/1/L2} {\bibfield  {journal} {\bibinfo
  {journal} {\apjl}\ }\textbf {\bibinfo {volume} {832}},\ \bibinfo {eid} {L2}
  (\bibinfo {year} {2016})},\ \Eprint {https://arxiv.org/abs/1609.05916}
  {arXiv:1609.05916 [astro-ph.HE]} \BibitemShut {NoStop}%
\bibitem [{\citenamefont {{Doctor}}\ \emph {et~al.}(2020)\citenamefont
  {{Doctor}}, \citenamefont {{Wysocki}}, \citenamefont {{O'Shaughnessy}},
  \citenamefont {{Holz}},\ and\ \citenamefont {{Farr}}}]{Doctor19}%
  \BibitemOpen
  \bibfield  {author} {\bibinfo {author} {\bibfnamefont {Z.}~\bibnamefont
  {{Doctor}}}, \bibinfo {author} {\bibfnamefont {D.}~\bibnamefont {{Wysocki}}},
  \bibinfo {author} {\bibfnamefont {R.}~\bibnamefont {{O'Shaughnessy}}},
  \bibinfo {author} {\bibfnamefont {D.~E.}\ \bibnamefont {{Holz}}},\ and\
  \bibinfo {author} {\bibfnamefont {B.}~\bibnamefont {{Farr}}},\ }\bibfield
  {title} {\bibinfo {title} {{Black Hole Coagulation: Modeling Hierarchical
  Mergers in Black Hole Populations}},\ }\href
  {https://doi.org/10.3847/1538-4357/ab7fac} {\bibfield  {journal} {\bibinfo
  {journal} {\apj}\ }\textbf {\bibinfo {volume} {893}},\ \bibinfo {eid} {35}
  (\bibinfo {year} {2020})},\ \Eprint {https://arxiv.org/abs/1911.04424}
  {arXiv:1911.04424 [astro-ph.HE]} \BibitemShut {NoStop}%
\bibitem [{\citenamefont {Roulet}\ \emph {et~al.}(2021)\citenamefont {Roulet},
  \citenamefont {Chia}, \citenamefont {Olsen}, \citenamefont {Dai},
  \citenamefont {Venumadhav}, \citenamefont {Zackay},\ and\ \citenamefont
  {Zaldarriaga}}]{Roulet}%
  \BibitemOpen
  \bibfield  {author} {\bibinfo {author} {\bibfnamefont {J.}~\bibnamefont
  {Roulet}}, \bibinfo {author} {\bibfnamefont {H.~S.}\ \bibnamefont {Chia}},
  \bibinfo {author} {\bibfnamefont {S.}~\bibnamefont {Olsen}}, \bibinfo
  {author} {\bibfnamefont {L.}~\bibnamefont {Dai}}, \bibinfo {author}
  {\bibfnamefont {T.}~\bibnamefont {Venumadhav}}, \bibinfo {author}
  {\bibfnamefont {B.}~\bibnamefont {Zackay}},\ and\ \bibinfo {author}
  {\bibfnamefont {M.}~\bibnamefont {Zaldarriaga}},\ }\bibfield  {title}
  {\bibinfo {title} {Distribution of effective spins and masses of binary black
  holes from the ligo and virgo o1--o3a observing runs},\ }\href
  {https://doi.org/10.1103/PhysRevD.104.083010} {\bibfield  {journal} {\bibinfo
   {journal} {Phys. Rev. D}\ }\textbf {\bibinfo {volume} {104}},\ \bibinfo
  {pages} {083010} (\bibinfo {year} {2021})}\BibitemShut {NoStop}%
\bibitem [{\citenamefont {{Galaudage}}\ \emph {et~al.}(2021)\citenamefont
  {{Galaudage}}, \citenamefont {{Talbot}}, \citenamefont {{Nagar}},
  \citenamefont {{Jain}}, \citenamefont {{Thrane}},\ and\ \citenamefont
  {{Mandel}}}]{Galaudage21}%
  \BibitemOpen
  \bibfield  {author} {\bibinfo {author} {\bibfnamefont {S.}~\bibnamefont
  {{Galaudage}}}, \bibinfo {author} {\bibfnamefont {C.}~\bibnamefont
  {{Talbot}}}, \bibinfo {author} {\bibfnamefont {T.}~\bibnamefont {{Nagar}}},
  \bibinfo {author} {\bibfnamefont {D.}~\bibnamefont {{Jain}}}, \bibinfo
  {author} {\bibfnamefont {E.}~\bibnamefont {{Thrane}}},\ and\ \bibinfo
  {author} {\bibfnamefont {I.}~\bibnamefont {{Mandel}}},\ }\bibfield  {title}
  {\bibinfo {title} {{Building Better Spin Models for Merging Binary Black
  Holes: Evidence for Nonspinning and Rapidly Spinning Nearly Aligned
  Subpopulations}},\ }\href {https://doi.org/10.3847/2041-8213/ac2f3c}
  {\bibfield  {journal} {\bibinfo  {journal} {\apjl}\ }\textbf {\bibinfo
  {volume} {921}},\ \bibinfo {eid} {L15} (\bibinfo {year} {2021})},\ \Eprint
  {https://arxiv.org/abs/2109.02424} {arXiv:2109.02424 [gr-qc]} \BibitemShut
  {NoStop}%
\bibitem [{\citenamefont {{Callister}}\ \emph {et~al.}(2022)\citenamefont
  {{Callister}}, \citenamefont {{Miller}}, \citenamefont {{Chatziioannou}},\
  and\ \citenamefont {{Farr}}}]{Callister22}%
  \BibitemOpen
  \bibfield  {author} {\bibinfo {author} {\bibfnamefont {T.~A.}\ \bibnamefont
  {{Callister}}}, \bibinfo {author} {\bibfnamefont {S.~J.}\ \bibnamefont
  {{Miller}}}, \bibinfo {author} {\bibfnamefont {K.}~\bibnamefont
  {{Chatziioannou}}},\ and\ \bibinfo {author} {\bibfnamefont {W.~M.}\
  \bibnamefont {{Farr}}},\ }\bibfield  {title} {\bibinfo {title} {{No evidence
  that the majority of black holes in binaries have zero spin}},\ }\href@noop
  {} {\bibfield  {journal} {\bibinfo  {journal} {arXiv e-prints}\ ,\ \bibinfo
  {eid} {arXiv:2205.08574}} (\bibinfo {year} {2022})},\ \Eprint
  {https://arxiv.org/abs/2205.08574} {arXiv:2205.08574 [astro-ph.HE]}
  \BibitemShut {NoStop}%
\bibitem [{\citenamefont {{Payne}}\ and\ \citenamefont
  {{Thrane}}(2022)}]{Payne22}%
  \BibitemOpen
  \bibfield  {author} {\bibinfo {author} {\bibfnamefont {E.}~\bibnamefont
  {{Payne}}}\ and\ \bibinfo {author} {\bibfnamefont {E.}~\bibnamefont
  {{Thrane}}},\ }\bibfield  {title} {\bibinfo {title} {{Model exploration in
  gravitational-wave astronomy with the maximum population likelihood}},\
  }\href@noop {} {\bibfield  {journal} {\bibinfo  {journal} {arXiv e-prints}\
  ,\ \bibinfo {eid} {arXiv:2210.11641}} (\bibinfo {year} {2022})},\ \Eprint
  {https://arxiv.org/abs/2210.11641} {arXiv:2210.11641 [astro-ph.IM]}
  \BibitemShut {NoStop}%
\bibitem [{\citenamefont {Romero-Shaw}\ \emph {et~al.}(2022)\citenamefont
  {Romero-Shaw}, \citenamefont {Thrane},\ and\ \citenamefont
  {Lasky}}]{RomeroShaw22}%
  \BibitemOpen
  \bibfield  {author} {\bibinfo {author} {\bibfnamefont {I.~M.}\ \bibnamefont
  {Romero-Shaw}}, \bibinfo {author} {\bibfnamefont {E.}~\bibnamefont
  {Thrane}},\ and\ \bibinfo {author} {\bibfnamefont {P.~D.}\ \bibnamefont
  {Lasky}},\ }\bibfield  {title} {\bibinfo {title} {When models fail: An
  introduction to posterior predictive checks and model misspecification in
  gravitational-wave astronomy},\ }\href {https://doi.org/10.1017/pasa.2022.24}
  {\bibfield  {journal} {\bibinfo  {journal} {Publications of the Astronomical
  Society of Australia}\ }\textbf {\bibinfo {volume} {39}},\ \bibinfo {pages}
  {e025} (\bibinfo {year} {2022})}\BibitemShut {NoStop}%
\bibitem [{\citenamefont {{Wysocki}}\ \emph {et~al.}(2019)\citenamefont
  {{Wysocki}}, \citenamefont {{Lange}},\ and\ \citenamefont
  {{O'Shaughnessy}}}]{Wysocki19}%
  \BibitemOpen
  \bibfield  {author} {\bibinfo {author} {\bibfnamefont {D.}~\bibnamefont
  {{Wysocki}}}, \bibinfo {author} {\bibfnamefont {J.}~\bibnamefont {{Lange}}},\
  and\ \bibinfo {author} {\bibfnamefont {R.}~\bibnamefont {{O'Shaughnessy}}},\
  }\bibfield  {title} {\bibinfo {title} {{Reconstructing phenomenological
  distributions of compact binaries via gravitational wave observations}},\
  }\href {https://doi.org/10.1103/PhysRevD.100.043012} {\bibfield  {journal}
  {\bibinfo  {journal} {\prd}\ }\textbf {\bibinfo {volume} {100}},\ \bibinfo
  {eid} {043012} (\bibinfo {year} {2019})},\ \Eprint
  {https://arxiv.org/abs/1805.06442} {arXiv:1805.06442 [gr-qc]} \BibitemShut
  {NoStop}%
\bibitem [{\citenamefont {{Talbot}}\ and\ \citenamefont
  {{Thrane}}(2017)}]{Talbotspin}%
  \BibitemOpen
  \bibfield  {author} {\bibinfo {author} {\bibfnamefont {C.}~\bibnamefont
  {{Talbot}}}\ and\ \bibinfo {author} {\bibfnamefont {E.}~\bibnamefont
  {{Thrane}}},\ }\bibfield  {title} {\bibinfo {title} {{Determining the
  population properties of spinning black holes}},\ }\href
  {https://doi.org/10.1103/PhysRevD.96.023012} {\bibfield  {journal} {\bibinfo
  {journal} {\prd}\ }\textbf {\bibinfo {volume} {96}},\ \bibinfo {eid} {023012}
  (\bibinfo {year} {2017})},\ \Eprint {https://arxiv.org/abs/1704.08370}
  {arXiv:1704.08370 [astro-ph.HE]} \BibitemShut {NoStop}%
\bibitem [{\citenamefont {{Vitale}}\ \emph
  {et~al.}(2022{\natexlab{a}})\citenamefont {{Vitale}}, \citenamefont
  {{Biscoveanu}},\ and\ \citenamefont {{Talbot}}}]{Vitale2022}%
  \BibitemOpen
  \bibfield  {author} {\bibinfo {author} {\bibfnamefont {S.}~\bibnamefont
  {{Vitale}}}, \bibinfo {author} {\bibfnamefont {S.}~\bibnamefont
  {{Biscoveanu}}},\ and\ \bibinfo {author} {\bibfnamefont {C.}~\bibnamefont
  {{Talbot}}},\ }\bibfield  {title} {\bibinfo {title} {{Spin it as you like:
  the (lack of a) measurement of the spin tilt distribution with
  LIGO-Virgo-KAGRA binary black holes}},\ }\href@noop {} {\bibfield  {journal}
  {\bibinfo  {journal} {arXiv e-prints}\ ,\ \bibinfo {eid} {arXiv:2209.06978}}
  (\bibinfo {year} {2022}{\natexlab{a}})},\ \Eprint
  {https://arxiv.org/abs/2209.06978} {arXiv:2209.06978 [astro-ph.HE]}
  \BibitemShut {NoStop}%
\bibitem [{\citenamefont {{Edelman}}\ \emph
  {et~al.}(2022{\natexlab{a}})\citenamefont {{Edelman}}, \citenamefont
  {{Doctor}}, \citenamefont {{Godfrey}},\ and\ \citenamefont
  {{Farr}}}]{Edelman}%
  \BibitemOpen
  \bibfield  {author} {\bibinfo {author} {\bibfnamefont {B.}~\bibnamefont
  {{Edelman}}}, \bibinfo {author} {\bibfnamefont {Z.}~\bibnamefont {{Doctor}}},
  \bibinfo {author} {\bibfnamefont {J.}~\bibnamefont {{Godfrey}}},\ and\
  \bibinfo {author} {\bibfnamefont {B.}~\bibnamefont {{Farr}}},\ }\bibfield
  {title} {\bibinfo {title} {{Ain't No Mountain High Enough: Semiparametric
  Modeling of LIGO-Virgo's Binary Black Hole Mass Distribution}},\ }\href
  {https://doi.org/10.3847/1538-4357/ac3667} {\bibfield  {journal} {\bibinfo
  {journal} {\apj}\ }\textbf {\bibinfo {volume} {924}},\ \bibinfo {eid} {101}
  (\bibinfo {year} {2022}{\natexlab{a}})},\ \Eprint
  {https://arxiv.org/abs/2109.06137} {arXiv:2109.06137 [astro-ph.HE]}
  \BibitemShut {NoStop}%
\bibitem [{\citenamefont {{Thrane}}\ and\ \citenamefont
  {{Talbot}}(2019)}]{thranetalbot19}%
  \BibitemOpen
  \bibfield  {author} {\bibinfo {author} {\bibfnamefont {E.}~\bibnamefont
  {{Thrane}}}\ and\ \bibinfo {author} {\bibfnamefont {C.}~\bibnamefont
  {{Talbot}}},\ }\bibfield  {title} {\bibinfo {title} {{An introduction to
  Bayesian inference in gravitational-wave astronomy: Parameter estimation,
  model selection, and hierarchical models}},\ }\href
  {https://doi.org/10.1017/pasa.2019.2} {\bibfield  {journal} {\bibinfo
  {journal} {\pasa}\ }\textbf {\bibinfo {volume} {36}},\ \bibinfo {eid} {e010}
  (\bibinfo {year} {2019})},\ \Eprint {https://arxiv.org/abs/1809.02293}
  {arXiv:1809.02293 [astro-ph.IM]} \BibitemShut {NoStop}%
\bibitem [{\citenamefont {{Talbot}}\ and\ \citenamefont
  {{Thrane}}(2018)}]{TalbotPowerLawPeak}%
  \BibitemOpen
  \bibfield  {author} {\bibinfo {author} {\bibfnamefont {C.}~\bibnamefont
  {{Talbot}}}\ and\ \bibinfo {author} {\bibfnamefont {E.}~\bibnamefont
  {{Thrane}}},\ }\bibfield  {title} {\bibinfo {title} {{Measuring the Binary
  Black Hole Mass Spectrum with an Astrophysically Motivated
  Parameterization}},\ }\href {https://doi.org/10.3847/1538-4357/aab34c}
  {\bibfield  {journal} {\bibinfo  {journal} {\apj}\ }\textbf {\bibinfo
  {volume} {856}},\ \bibinfo {eid} {173} (\bibinfo {year} {2018})},\ \Eprint
  {https://arxiv.org/abs/1801.02699} {arXiv:1801.02699 [astro-ph.HE]}
  \BibitemShut {NoStop}%
\bibitem [{\citenamefont {{Fishbach}}\ \emph {et~al.}(2018)\citenamefont
  {{Fishbach}}, \citenamefont {{Holz}},\ and\ \citenamefont
  {{Farr}}}]{Fishbach18}%
  \BibitemOpen
  \bibfield  {author} {\bibinfo {author} {\bibfnamefont {M.}~\bibnamefont
  {{Fishbach}}}, \bibinfo {author} {\bibfnamefont {D.~E.}\ \bibnamefont
  {{Holz}}},\ and\ \bibinfo {author} {\bibfnamefont {W.~M.}\ \bibnamefont
  {{Farr}}},\ }\bibfield  {title} {\bibinfo {title} {{Does the Black Hole
  Merger Rate Evolve with Redshift?}},\ }\href
  {https://doi.org/10.3847/2041-8213/aad800} {\bibfield  {journal} {\bibinfo
  {journal} {\apjl}\ }\textbf {\bibinfo {volume} {863}},\ \bibinfo {eid} {L41}
  (\bibinfo {year} {2018})},\ \Eprint {https://arxiv.org/abs/1805.10270}
  {arXiv:1805.10270 [astro-ph.HE]} \BibitemShut {NoStop}%
\bibitem [{\citenamefont {{The LIGO Scientific Collaboration}}\ and\
  \citenamefont {{the Virgo Collaboration}}(2019)}]{gwtc1pops}%
  \BibitemOpen
  \bibfield  {author} {\bibinfo {author} {\bibnamefont {{The LIGO Scientific
  Collaboration}}}\ and\ \bibinfo {author} {\bibnamefont {{the Virgo
  Collaboration}}},\ }\bibfield  {title} {\bibinfo {title} {{Binary Black Hole
  Population Properties Inferred from the First and Second Observing Runs of
  Advanced LIGO and Advanced Virgo}},\ }\href
  {https://doi.org/10.3847/2041-8213/ab3800} {\bibfield  {journal} {\bibinfo
  {journal} {\apjl}\ }\textbf {\bibinfo {volume} {882}},\ \bibinfo {eid} {L24}
  (\bibinfo {year} {2019})},\ \Eprint {https://arxiv.org/abs/1811.12940}
  {arXiv:1811.12940 [astro-ph.HE]} \BibitemShut {NoStop}%
\bibitem [{\citenamefont {Talbot}\ \emph {et~al.}(2019)\citenamefont {Talbot},
  \citenamefont {Smith}, \citenamefont {Thrane},\ and\ \citenamefont
  {Poole}}]{Talbotgwpop}%
  \BibitemOpen
  \bibfield  {author} {\bibinfo {author} {\bibfnamefont {C.}~\bibnamefont
  {Talbot}}, \bibinfo {author} {\bibfnamefont {R.}~\bibnamefont {Smith}},
  \bibinfo {author} {\bibfnamefont {E.}~\bibnamefont {Thrane}},\ and\ \bibinfo
  {author} {\bibfnamefont {G.~B.}\ \bibnamefont {Poole}},\ }\bibfield  {title}
  {\bibinfo {title} {Parallelized inference for gravitational-wave astronomy},\
  }\href {https://doi.org/10.1103/PhysRevD.100.043030} {\bibfield  {journal}
  {\bibinfo  {journal} {Phys. Rev. D}\ }\textbf {\bibinfo {volume} {100}},\
  \bibinfo {pages} {043030} (\bibinfo {year} {2019})}\BibitemShut {NoStop}%
\bibitem [{\citenamefont {{Speagle}}(2020)}]{Speagledynesty}%
  \BibitemOpen
  \bibfield  {author} {\bibinfo {author} {\bibfnamefont {J.~S.}\ \bibnamefont
  {{Speagle}}},\ }\bibfield  {title} {\bibinfo {title} {{DYNESTY: a dynamic
  nested sampling package for estimating Bayesian posteriors and evidences}},\
  }\href {https://doi.org/10.1093/mnras/staa278} {\bibfield  {journal}
  {\bibinfo  {journal} {\mnras}\ }\textbf {\bibinfo {volume} {493}},\ \bibinfo
  {pages} {3132} (\bibinfo {year} {2020})},\ \Eprint
  {https://arxiv.org/abs/1904.02180} {arXiv:1904.02180 [astro-ph.IM]}
  \BibitemShut {NoStop}%
\bibitem [{\citenamefont {{Vitale}}\ \emph
  {et~al.}(2022{\natexlab{b}})\citenamefont {{Vitale}}, \citenamefont
  {{Gerosa}}, \citenamefont {{Farr}},\ and\ \citenamefont
  {{Taylor}}}]{Vitale2022b}%
  \BibitemOpen
  \bibfield  {author} {\bibinfo {author} {\bibfnamefont {S.}~\bibnamefont
  {{Vitale}}}, \bibinfo {author} {\bibfnamefont {D.}~\bibnamefont {{Gerosa}}},
  \bibinfo {author} {\bibfnamefont {W.~M.}\ \bibnamefont {{Farr}}},\ and\
  \bibinfo {author} {\bibfnamefont {S.~R.}\ \bibnamefont {{Taylor}}},\
  }\bibfield  {title} {\bibinfo {title} {{Inferring the Properties of a
  Population of Compact Binaries in Presence of Selection Effects}},\ }in\
  \href {https://doi.org/10.1007/978-981-15-4702-7_45-1} {\emph {\bibinfo
  {booktitle} {Handbook of Gravitational Wave Astronomy. Edited by C. Bambi}}}\
  (\bibinfo {year} {2022})\ p.~\bibinfo {pages} {45}\BibitemShut {NoStop}%
\bibitem [{\citenamefont {{LIGO Scientific Collaboration}}\ \emph
  {et~al.}(2016)\citenamefont {{LIGO Scientific Collaboration}}, \citenamefont
  {{Virgo Collaboration}} \emph {et~al.}}]{LIGO_O1}%
  \BibitemOpen
  \bibfield  {author} {\bibinfo {author} {\bibnamefont {{LIGO Scientific
  Collaboration}}}, \bibinfo {author} {\bibnamefont {{Virgo Collaboration}}},
  \emph {et~al.},\ }\bibfield  {title} {\bibinfo {title} {{Binary Black Hole
  Mergers in the First Advanced LIGO Observing Run}},\ }\href
  {https://doi.org/10.1103/PhysRevX.6.041015} {\bibfield  {journal} {\bibinfo
  {journal} {Physical Review X}\ }\textbf {\bibinfo {volume} {6}},\ \bibinfo
  {eid} {041015} (\bibinfo {year} {2016})},\ \Eprint
  {https://arxiv.org/abs/1606.04856} {arXiv:1606.04856 [gr-qc]} \BibitemShut
  {NoStop}%
\bibitem [{\citenamefont {{Farr}}(2019)}]{Farr19}%
  \BibitemOpen
  \bibfield  {author} {\bibinfo {author} {\bibfnamefont {W.~M.}\ \bibnamefont
  {{Farr}}},\ }\bibfield  {title} {\bibinfo {title} {{Accuracy Requirements for
  Empirically Measured Selection Functions}},\ }\href
  {https://doi.org/10.3847/2515-5172/ab1d5f} {\bibfield  {journal} {\bibinfo
  {journal} {Research Notes of the American Astronomical Society}\ }\textbf
  {\bibinfo {volume} {3}},\ \bibinfo {eid} {66} (\bibinfo {year} {2019})},\
  \Eprint {https://arxiv.org/abs/1904.10879} {arXiv:1904.10879 [astro-ph.IM]}
  \BibitemShut {NoStop}%
\bibitem [{\citenamefont {{LIGO Scientific Collaboration}}\ \emph
  {et~al.}(2021{\natexlab{b}})\citenamefont {{LIGO Scientific Collaboration}},
  \citenamefont {{Virgo Collaboration}} \emph {et~al.}}]{gwtc2}%
  \BibitemOpen
  \bibfield  {author} {\bibinfo {author} {\bibnamefont {{LIGO Scientific
  Collaboration}}}, \bibinfo {author} {\bibnamefont {{Virgo Collaboration}}},
  \emph {et~al.},\ }\bibfield  {title} {\bibinfo {title} {{GWTC-2: Compact
  Binary Coalescences Observed by LIGO and Virgo during the First Half of the
  Third Observing Run}},\ }\href {https://doi.org/10.1103/PhysRevX.11.021053}
  {\bibfield  {journal} {\bibinfo  {journal} {Physical Review X}\ }\textbf
  {\bibinfo {volume} {11}},\ \bibinfo {eid} {021053} (\bibinfo {year}
  {2021}{\natexlab{b}})},\ \Eprint {https://arxiv.org/abs/2010.14527}
  {arXiv:2010.14527 [gr-qc]} \BibitemShut {NoStop}%
\bibitem [{\citenamefont {{Golomb}}\ and\ \citenamefont
  {{Talbot}}(2022)}]{GolombTalbotbns}%
  \BibitemOpen
  \bibfield  {author} {\bibinfo {author} {\bibfnamefont {J.}~\bibnamefont
  {{Golomb}}}\ and\ \bibinfo {author} {\bibfnamefont {C.}~\bibnamefont
  {{Talbot}}},\ }\bibfield  {title} {\bibinfo {title} {{Hierarchical Inference
  of Binary Neutron Star Mass Distribution and Equation of State with
  Gravitational Waves}},\ }\href {https://doi.org/10.3847/1538-4357/ac43bc}
  {\bibfield  {journal} {\bibinfo  {journal} {\apj}\ }\textbf {\bibinfo
  {volume} {926}},\ \bibinfo {eid} {79} (\bibinfo {year} {2022})},\ \Eprint
  {https://arxiv.org/abs/2106.15745} {arXiv:2106.15745 [astro-ph.HE]}
  \BibitemShut {NoStop}%
\bibitem [{\citenamefont {{Martino}}\ \emph {et~al.}(2016)\citenamefont
  {{Martino}}, \citenamefont {{Elvira}},\ and\ \citenamefont
  {{Louzada}}}]{montecarlo}%
  \BibitemOpen
  \bibfield  {author} {\bibinfo {author} {\bibfnamefont {L.}~\bibnamefont
  {{Martino}}}, \bibinfo {author} {\bibfnamefont {V.}~\bibnamefont
  {{Elvira}}},\ and\ \bibinfo {author} {\bibfnamefont {F.}~\bibnamefont
  {{Louzada}}},\ }\bibfield  {title} {\bibinfo {title} {{Effective Sample Size
  for Importance Sampling based on discrepancy measures}},\ }\href@noop {}
  {\bibfield  {journal} {\bibinfo  {journal} {arXiv e-prints}\ ,\ \bibinfo
  {eid} {arXiv:1602.03572}} (\bibinfo {year} {2016})},\ \Eprint
  {https://arxiv.org/abs/1602.03572} {arXiv:1602.03572 [stat.CO]} \BibitemShut
  {NoStop}%
\bibitem [{\citenamefont {{Essick}}\ and\ \citenamefont
  {{Farr}}(2022)}]{Essick22}%
  \BibitemOpen
  \bibfield  {author} {\bibinfo {author} {\bibfnamefont {R.}~\bibnamefont
  {{Essick}}}\ and\ \bibinfo {author} {\bibfnamefont {W.}~\bibnamefont
  {{Farr}}},\ }\bibfield  {title} {\bibinfo {title} {{Precision Requirements
  for Monte Carlo Sums within Hierarchical Bayesian Inference}},\ }\href@noop
  {} {\bibfield  {journal} {\bibinfo  {journal} {arXiv e-prints}\ ,\ \bibinfo
  {eid} {arXiv:2204.00461}} (\bibinfo {year} {2022})},\ \Eprint
  {https://arxiv.org/abs/2204.00461} {arXiv:2204.00461 [astro-ph.IM]}
  \BibitemShut {NoStop}%
\bibitem [{Note1()}]{Note1}%
  \BibitemOpen
  \bibinfo {note} {Our spline model does not support $p = 0$ for any point in
  parameter space, as the spline is exponentiated, but the prior does allow for
  arbitrarily low values and we would therefore expect that the posterior for
  $p(a = 0)$ would approach 0 if the data truly favored the non-presence of
  zero-spin BHs.}\BibitemShut {Stop}%
\bibitem [{\citenamefont {{Schmidt}}\ \emph {et~al.}(2015)\citenamefont
  {{Schmidt}}, \citenamefont {{Ohme}},\ and\ \citenamefont
  {{Hannam}}}]{Schmidt15}%
  \BibitemOpen
  \bibfield  {author} {\bibinfo {author} {\bibfnamefont {P.}~\bibnamefont
  {{Schmidt}}}, \bibinfo {author} {\bibfnamefont {F.}~\bibnamefont {{Ohme}}},\
  and\ \bibinfo {author} {\bibfnamefont {M.}~\bibnamefont {{Hannam}}},\
  }\bibfield  {title} {\bibinfo {title} {{Towards models of gravitational
  waveforms from generic binaries: II. Modelling precession effects with a
  single effective precession parameter}},\ }\href
  {https://doi.org/10.1103/PhysRevD.91.024043} {\bibfield  {journal} {\bibinfo
  {journal} {\prd}\ }\textbf {\bibinfo {volume} {91}},\ \bibinfo {eid} {024043}
  (\bibinfo {year} {2015})},\ \Eprint {https://arxiv.org/abs/1408.1810}
  {arXiv:1408.1810 [gr-qc]} \BibitemShut {NoStop}%
\bibitem [{\citenamefont {{Miller}}\ \emph {et~al.}(2020)\citenamefont
  {{Miller}}, \citenamefont {{Callister}},\ and\ \citenamefont
  {{Farr}}}]{Miller2020}%
  \BibitemOpen
  \bibfield  {author} {\bibinfo {author} {\bibfnamefont {S.}~\bibnamefont
  {{Miller}}}, \bibinfo {author} {\bibfnamefont {T.~A.}\ \bibnamefont
  {{Callister}}},\ and\ \bibinfo {author} {\bibfnamefont {W.~M.}\ \bibnamefont
  {{Farr}}},\ }\bibfield  {title} {\bibinfo {title} {{The Low Effective Spin of
  Binary Black Holes and Implications for Individual Gravitational-wave
  Events}},\ }\href {https://doi.org/10.3847/1538-4357/ab80c0} {\bibfield
  {journal} {\bibinfo  {journal} {\apj}\ }\textbf {\bibinfo {volume} {895}},\
  \bibinfo {eid} {128} (\bibinfo {year} {2020})},\ \Eprint
  {https://arxiv.org/abs/2001.06051} {arXiv:2001.06051 [astro-ph.HE]}
  \BibitemShut {NoStop}%
\bibitem [{\citenamefont {{Zevin}}\ \emph {et~al.}(2021)\citenamefont
  {{Zevin}}, \citenamefont {{Bavera}}, \citenamefont {{Berry}}, \citenamefont
  {{Kalogera}}, \citenamefont {{Fragos}}, \citenamefont {{Marchant}},
  \citenamefont {{Rodriguez}}, \citenamefont {{Antonini}}, \citenamefont
  {{Holz}},\ and\ \citenamefont {{Pankow}}}]{Zevin2021}%
  \BibitemOpen
  \bibfield  {author} {\bibinfo {author} {\bibfnamefont {M.}~\bibnamefont
  {{Zevin}}}, \bibinfo {author} {\bibfnamefont {S.~S.}\ \bibnamefont
  {{Bavera}}}, \bibinfo {author} {\bibfnamefont {C.~P.~L.}\ \bibnamefont
  {{Berry}}}, \bibinfo {author} {\bibfnamefont {V.}~\bibnamefont {{Kalogera}}},
  \bibinfo {author} {\bibfnamefont {T.}~\bibnamefont {{Fragos}}}, \bibinfo
  {author} {\bibfnamefont {P.}~\bibnamefont {{Marchant}}}, \bibinfo {author}
  {\bibfnamefont {C.~L.}\ \bibnamefont {{Rodriguez}}}, \bibinfo {author}
  {\bibfnamefont {F.}~\bibnamefont {{Antonini}}}, \bibinfo {author}
  {\bibfnamefont {D.~E.}\ \bibnamefont {{Holz}}},\ and\ \bibinfo {author}
  {\bibfnamefont {C.}~\bibnamefont {{Pankow}}},\ }\bibfield  {title} {\bibinfo
  {title} {{One Channel to Rule Them All? Constraining the Origins of Binary
  Black Holes Using Multiple Formation Pathways}},\ }\href
  {https://doi.org/10.3847/1538-4357/abe40e} {\bibfield  {journal} {\bibinfo
  {journal} {\apj}\ }\textbf {\bibinfo {volume} {910}},\ \bibinfo {eid} {152}
  (\bibinfo {year} {2021})},\ \Eprint {https://arxiv.org/abs/2011.10057}
  {arXiv:2011.10057 [astro-ph.HE]} \BibitemShut {NoStop}%
\bibitem [{\citenamefont {{Wong}}\ \emph {et~al.}(2021)\citenamefont {{Wong}},
  \citenamefont {{Breivik}}, \citenamefont {{Kremer}},\ and\ \citenamefont
  {{Callister}}}]{Wong2021}%
  \BibitemOpen
  \bibfield  {author} {\bibinfo {author} {\bibfnamefont {K.~W.~K.}\
  \bibnamefont {{Wong}}}, \bibinfo {author} {\bibfnamefont {K.}~\bibnamefont
  {{Breivik}}}, \bibinfo {author} {\bibfnamefont {K.}~\bibnamefont
  {{Kremer}}},\ and\ \bibinfo {author} {\bibfnamefont {T.}~\bibnamefont
  {{Callister}}},\ }\bibfield  {title} {\bibinfo {title} {{Joint constraints on
  the field-cluster mixing fraction, common envelope efficiency, and globular
  cluster radii from a population of binary hole mergers via deep learning}},\
  }\href {https://doi.org/10.1103/PhysRevD.103.083021} {\bibfield  {journal}
  {\bibinfo  {journal} {\prd}\ }\textbf {\bibinfo {volume} {103}},\ \bibinfo
  {eid} {083021} (\bibinfo {year} {2021})},\ \Eprint
  {https://arxiv.org/abs/2011.03564} {arXiv:2011.03564 [astro-ph.HE]}
  \BibitemShut {NoStop}%
\bibitem [{\citenamefont {{Mandel}}\ \emph {et~al.}(2017)\citenamefont
  {{Mandel}}, \citenamefont {{Farr}}, \citenamefont {{Colonna}}, \citenamefont
  {{Stevenson}}, \citenamefont {{Ti{\v{n}}o}},\ and\ \citenamefont
  {{Veitch}}}]{Mandel2017}%
  \BibitemOpen
  \bibfield  {author} {\bibinfo {author} {\bibfnamefont {I.}~\bibnamefont
  {{Mandel}}}, \bibinfo {author} {\bibfnamefont {W.~M.}\ \bibnamefont
  {{Farr}}}, \bibinfo {author} {\bibfnamefont {A.}~\bibnamefont {{Colonna}}},
  \bibinfo {author} {\bibfnamefont {S.}~\bibnamefont {{Stevenson}}}, \bibinfo
  {author} {\bibfnamefont {P.}~\bibnamefont {{Ti{\v{n}}o}}},\ and\ \bibinfo
  {author} {\bibfnamefont {J.}~\bibnamefont {{Veitch}}},\ }\bibfield  {title}
  {\bibinfo {title} {{Model-independent inference on compact-binary
  observations}},\ }\href {https://doi.org/10.1093/mnras/stw2883} {\bibfield
  {journal} {\bibinfo  {journal} {\mnras}\ }\textbf {\bibinfo {volume} {465}},\
  \bibinfo {pages} {3254} (\bibinfo {year} {2017})},\ \Eprint
  {https://arxiv.org/abs/1608.08223} {arXiv:1608.08223 [astro-ph.HE]}
  \BibitemShut {NoStop}%
\bibitem [{\citenamefont {{Tiwari}}(2021)}]{Tiwari2021}%
  \BibitemOpen
  \bibfield  {author} {\bibinfo {author} {\bibfnamefont {V.}~\bibnamefont
  {{Tiwari}}},\ }\bibfield  {title} {\bibinfo {title} {{VAMANA: modeling binary
  black hole population with minimal assumptions}},\ }\href
  {https://doi.org/10.1088/1361-6382/ac0b54} {\bibfield  {journal} {\bibinfo
  {journal} {Classical and Quantum Gravity}\ }\textbf {\bibinfo {volume}
  {38}},\ \bibinfo {eid} {155007} (\bibinfo {year} {2021})},\ \Eprint
  {https://arxiv.org/abs/2006.15047} {arXiv:2006.15047 [astro-ph.HE]}
  \BibitemShut {NoStop}%
\bibitem [{\citenamefont {{Rinaldi}}\ and\ \citenamefont {{Del
  Pozzo}}(2022)}]{Rinaldi2022}%
  \BibitemOpen
  \bibfield  {author} {\bibinfo {author} {\bibfnamefont {S.}~\bibnamefont
  {{Rinaldi}}}\ and\ \bibinfo {author} {\bibfnamefont {W.}~\bibnamefont {{Del
  Pozzo}}},\ }\bibfield  {title} {\bibinfo {title} {{(H)DPGMM: a hierarchy of
  Dirichlet process Gaussian mixture models for the inference of the black hole
  mass function}},\ }\href {https://doi.org/10.1093/mnras/stab3224} {\bibfield
  {journal} {\bibinfo  {journal} {\mnras}\ }\textbf {\bibinfo {volume} {509}},\
  \bibinfo {pages} {5454} (\bibinfo {year} {2022})},\ \Eprint
  {https://arxiv.org/abs/2109.05960} {arXiv:2109.05960 [astro-ph.IM]}
  \BibitemShut {NoStop}%
\bibitem [{\citenamefont {{Belczynski}}\ \emph {et~al.}(2020)\citenamefont
  {{Belczynski}}, \citenamefont {{Klencki}}, \citenamefont {{Fields}},
  \citenamefont {{Olejak}}, \citenamefont {{Berti}}, \citenamefont {{Meynet}},
  \citenamefont {{Fryer}}, \citenamefont {{Holz}}, \citenamefont
  {{O'Shaughnessy}}, \citenamefont {{Brown}}, \citenamefont {{Bulik}},
  \citenamefont {{Leung}}, \citenamefont {{Nomoto}}, \citenamefont {{Madau}},
  \citenamefont {{Hirschi}}, \citenamefont {{Kaiser}}, \citenamefont {{Jones}},
  \citenamefont {{Mondal}}, \citenamefont {{Chruslinska}}, \citenamefont
  {{Drozda}}, \citenamefont {{Gerosa}}, \citenamefont {{Doctor}}, \citenamefont
  {{Giersz}}, \citenamefont {{Ekstrom}}, \citenamefont {{Georgy}},
  \citenamefont {{Askar}}, \citenamefont {{Baibhav}}, \citenamefont
  {{Wysocki}}, \citenamefont {{Natan}}, \citenamefont {{Farr}}, \citenamefont
  {{Wiktorowicz}}, \citenamefont {{Coleman Miller}}, \citenamefont {{Farr}},\
  and\ \citenamefont {{Lasota}}}]{Belczynski20}%
  \BibitemOpen
  \bibfield  {author} {\bibinfo {author} {\bibfnamefont {K.}~\bibnamefont
  {{Belczynski}}}, \bibinfo {author} {\bibfnamefont {J.}~\bibnamefont
  {{Klencki}}}, \bibinfo {author} {\bibfnamefont {C.~E.}\ \bibnamefont
  {{Fields}}}, \bibinfo {author} {\bibfnamefont {A.}~\bibnamefont {{Olejak}}},
  \bibinfo {author} {\bibfnamefont {E.}~\bibnamefont {{Berti}}}, \bibinfo
  {author} {\bibfnamefont {G.}~\bibnamefont {{Meynet}}}, \bibinfo {author}
  {\bibfnamefont {C.~L.}\ \bibnamefont {{Fryer}}}, \bibinfo {author}
  {\bibfnamefont {D.~E.}\ \bibnamefont {{Holz}}}, \bibinfo {author}
  {\bibfnamefont {R.}~\bibnamefont {{O'Shaughnessy}}}, \bibinfo {author}
  {\bibfnamefont {D.~A.}\ \bibnamefont {{Brown}}}, \bibinfo {author}
  {\bibfnamefont {T.}~\bibnamefont {{Bulik}}}, \bibinfo {author} {\bibfnamefont
  {S.~C.}\ \bibnamefont {{Leung}}}, \bibinfo {author} {\bibfnamefont
  {K.}~\bibnamefont {{Nomoto}}}, \bibinfo {author} {\bibfnamefont
  {P.}~\bibnamefont {{Madau}}}, \bibinfo {author} {\bibfnamefont
  {R.}~\bibnamefont {{Hirschi}}}, \bibinfo {author} {\bibfnamefont
  {E.}~\bibnamefont {{Kaiser}}}, \bibinfo {author} {\bibfnamefont
  {S.}~\bibnamefont {{Jones}}}, \bibinfo {author} {\bibfnamefont
  {S.}~\bibnamefont {{Mondal}}}, \bibinfo {author} {\bibfnamefont
  {M.}~\bibnamefont {{Chruslinska}}}, \bibinfo {author} {\bibfnamefont
  {P.}~\bibnamefont {{Drozda}}}, \bibinfo {author} {\bibfnamefont
  {D.}~\bibnamefont {{Gerosa}}}, \bibinfo {author} {\bibfnamefont
  {Z.}~\bibnamefont {{Doctor}}}, \bibinfo {author} {\bibfnamefont
  {M.}~\bibnamefont {{Giersz}}}, \bibinfo {author} {\bibfnamefont
  {S.}~\bibnamefont {{Ekstrom}}}, \bibinfo {author} {\bibfnamefont
  {C.}~\bibnamefont {{Georgy}}}, \bibinfo {author} {\bibfnamefont
  {A.}~\bibnamefont {{Askar}}}, \bibinfo {author} {\bibfnamefont
  {V.}~\bibnamefont {{Baibhav}}}, \bibinfo {author} {\bibfnamefont
  {D.}~\bibnamefont {{Wysocki}}}, \bibinfo {author} {\bibfnamefont
  {T.}~\bibnamefont {{Natan}}}, \bibinfo {author} {\bibfnamefont {W.~M.}\
  \bibnamefont {{Farr}}}, \bibinfo {author} {\bibfnamefont {G.}~\bibnamefont
  {{Wiktorowicz}}}, \bibinfo {author} {\bibfnamefont {M.}~\bibnamefont
  {{Coleman Miller}}}, \bibinfo {author} {\bibfnamefont {B.}~\bibnamefont
  {{Farr}}},\ and\ \bibinfo {author} {\bibfnamefont {J.~P.}\ \bibnamefont
  {{Lasota}}},\ }\bibfield  {title} {\bibinfo {title} {{Evolutionary roads
  leading to low effective spins, high black hole masses, and O1/O2 rates for
  LIGO/Virgo binary black holes}},\ }\href
  {https://doi.org/10.1051/0004-6361/201936528} {\bibfield  {journal} {\bibinfo
   {journal} {\aap}\ }\textbf {\bibinfo {volume} {636}},\ \bibinfo {eid} {A104}
  (\bibinfo {year} {2020})},\ \Eprint {https://arxiv.org/abs/1706.07053}
  {arXiv:1706.07053 [astro-ph.HE]} \BibitemShut {NoStop}%
\bibitem [{\citenamefont {{Tong}}\ \emph {et~al.}(2022)\citenamefont {{Tong}},
  \citenamefont {{Galaudage}},\ and\ \citenamefont {{Thrane}}}]{Tong2022}%
  \BibitemOpen
  \bibfield  {author} {\bibinfo {author} {\bibfnamefont {H.}~\bibnamefont
  {{Tong}}}, \bibinfo {author} {\bibfnamefont {S.}~\bibnamefont
  {{Galaudage}}},\ and\ \bibinfo {author} {\bibfnamefont {E.}~\bibnamefont
  {{Thrane}}},\ }\bibfield  {title} {\bibinfo {title} {{The population
  properties of spinning black holes using Gravitational-wave Transient Catalog
  3}},\ }\href@noop {} {\bibfield  {journal} {\bibinfo  {journal} {arXiv
  e-prints}\ ,\ \bibinfo {eid} {arXiv:2209.02206}} (\bibinfo {year} {2022})},\
  \Eprint {https://arxiv.org/abs/2209.02206} {arXiv:2209.02206 [astro-ph.HE]}
  \BibitemShut {NoStop}%
\bibitem [{\citenamefont {{Talbot}}\ and\ \citenamefont
  {{Thrane}}(2020)}]{Talbot20}%
  \BibitemOpen
  \bibfield  {author} {\bibinfo {author} {\bibfnamefont {C.}~\bibnamefont
  {{Talbot}}}\ and\ \bibinfo {author} {\bibfnamefont {E.}~\bibnamefont
  {{Thrane}}},\ }\bibfield  {title} {\bibinfo {title} {{Fast, flexible, and
  accurate evaluation of gravitational-wave Malmquist bias with machine
  learning}},\ }\href@noop {} {\bibfield  {journal} {\bibinfo  {journal} {arXiv
  e-prints}\ ,\ \bibinfo {eid} {arXiv:2012.01317}} (\bibinfo {year} {2020})},\
  \Eprint {https://arxiv.org/abs/2012.01317} {arXiv:2012.01317 [gr-qc]}
  \BibitemShut {NoStop}%
\bibitem [{\citenamefont {{Wang}}\ \emph {et~al.}(2022)\citenamefont {{Wang}},
  \citenamefont {{Li}}, \citenamefont {{Vink}}, \citenamefont {{Fan}},
  \citenamefont {{Tang}}, \citenamefont {{Qin}},\ and\ \citenamefont
  {{Wei}}}]{Wang22}%
  \BibitemOpen
  \bibfield  {author} {\bibinfo {author} {\bibfnamefont {Y.-Z.}\ \bibnamefont
  {{Wang}}}, \bibinfo {author} {\bibfnamefont {Y.-J.}\ \bibnamefont {{Li}}},
  \bibinfo {author} {\bibfnamefont {J.~S.}\ \bibnamefont {{Vink}}}, \bibinfo
  {author} {\bibfnamefont {Y.-Z.}\ \bibnamefont {{Fan}}}, \bibinfo {author}
  {\bibfnamefont {S.-P.}\ \bibnamefont {{Tang}}}, \bibinfo {author}
  {\bibfnamefont {Y.}~\bibnamefont {{Qin}}},\ and\ \bibinfo {author}
  {\bibfnamefont {D.-M.}\ \bibnamefont {{Wei}}},\ }\bibfield  {title} {\bibinfo
  {title} {{Tight Constraint on the Maximum Mass of Stellar-origin Binary Black
  Holes and Evidence for Hierarchical Mergers in Gravitational Wave
  Observations}},\ }\href@noop {} {\bibfield  {journal} {\bibinfo  {journal}
  {arXiv e-prints}\ ,\ \bibinfo {eid} {arXiv:2208.11871}} (\bibinfo {year}
  {2022})},\ \Eprint {https://arxiv.org/abs/2208.11871} {arXiv:2208.11871
  [astro-ph.HE]} \BibitemShut {NoStop}%
\bibitem [{\citenamefont {{Kimball}}\ \emph {et~al.}(2020)\citenamefont
  {{Kimball}}, \citenamefont {{Talbot}}, \citenamefont {{Berry}}, \citenamefont
  {{Carney}}, \citenamefont {{Zevin}}, \citenamefont {{Thrane}},\ and\
  \citenamefont {{Kalogera}}}]{Kimball20}%
  \BibitemOpen
  \bibfield  {author} {\bibinfo {author} {\bibfnamefont {C.}~\bibnamefont
  {{Kimball}}}, \bibinfo {author} {\bibfnamefont {C.}~\bibnamefont {{Talbot}}},
  \bibinfo {author} {\bibfnamefont {C.~P.~L.}\ \bibnamefont {{Berry}}},
  \bibinfo {author} {\bibfnamefont {M.}~\bibnamefont {{Carney}}}, \bibinfo
  {author} {\bibfnamefont {M.}~\bibnamefont {{Zevin}}}, \bibinfo {author}
  {\bibfnamefont {E.}~\bibnamefont {{Thrane}}},\ and\ \bibinfo {author}
  {\bibfnamefont {V.}~\bibnamefont {{Kalogera}}},\ }\bibfield  {title}
  {\bibinfo {title} {{Black Hole Genealogy: Identifying Hierarchical Mergers
  with Gravitational Waves}},\ }\href
  {https://doi.org/10.3847/1538-4357/aba518} {\bibfield  {journal} {\bibinfo
  {journal} {\apj}\ }\textbf {\bibinfo {volume} {900}},\ \bibinfo {eid} {177}
  (\bibinfo {year} {2020})},\ \Eprint {https://arxiv.org/abs/2005.00023}
  {arXiv:2005.00023 [astro-ph.HE]} \BibitemShut {NoStop}%
\bibitem [{\citenamefont {{Mould}}\ \emph {et~al.}(2022)\citenamefont
  {{Mould}}, \citenamefont {{Gerosa}}, \citenamefont {{Broekgaarden}},\ and\
  \citenamefont {{Steinle}}}]{Mould2022}%
  \BibitemOpen
  \bibfield  {author} {\bibinfo {author} {\bibfnamefont {M.}~\bibnamefont
  {{Mould}}}, \bibinfo {author} {\bibfnamefont {D.}~\bibnamefont {{Gerosa}}},
  \bibinfo {author} {\bibfnamefont {F.~S.}\ \bibnamefont {{Broekgaarden}}},\
  and\ \bibinfo {author} {\bibfnamefont {N.}~\bibnamefont {{Steinle}}},\
  }\bibfield  {title} {\bibinfo {title} {{Which black hole formed first?
  Mass-ratio reversal in massive binary stars from gravitational-wave data}},\
  }\bibfield  {journal} {\bibinfo  {journal} {\mnras}\ }\href
  {https://doi.org/10.1093/mnras/stac2859} {10.1093/mnras/stac2859} (\bibinfo
  {year} {2022}),\ \Eprint {https://arxiv.org/abs/2205.12329} {arXiv:2205.12329
  [astro-ph.HE]} \BibitemShut {NoStop}%
\bibitem [{\citenamefont {{Edelman}}\ \emph
  {et~al.}(2022{\natexlab{b}})\citenamefont {{Edelman}}, \citenamefont
  {{Farr}},\ and\ \citenamefont {{Doctor}}}]{Edelman22}%
  \BibitemOpen
  \bibfield  {author} {\bibinfo {author} {\bibfnamefont {B.}~\bibnamefont
  {{Edelman}}}, \bibinfo {author} {\bibfnamefont {B.}~\bibnamefont {{Farr}}},\
  and\ \bibinfo {author} {\bibfnamefont {Z.}~\bibnamefont {{Doctor}}},\
  }\bibfield  {title} {\bibinfo {title} {{Cover Your Basis: Comprehensive
  Data-Driven Characterization of the Binary Black Hole Population}},\
  }\href@noop {} {\bibfield  {journal} {\bibinfo  {journal} {arXiv e-prints}\
  ,\ \bibinfo {eid} {arXiv:2210.12834}} (\bibinfo {year}
  {2022}{\natexlab{b}})},\ \Eprint {https://arxiv.org/abs/2210.12834}
  {arXiv:2210.12834 [astro-ph.HE]} \BibitemShut {NoStop}%
\bibitem [{\citenamefont {{The LIGO Scientific Collaboration}}\ \emph
  {et~al.}(2021{\natexlab{e}})\citenamefont {{The LIGO Scientific
  Collaboration}}, \citenamefont {{the Virgo Collaboration}},\ and\
  \citenamefont {{the KAGRA Collaboration}}}]{lvko1o2o3injections}%
  \BibitemOpen
  \bibfield  {author} {\bibinfo {author} {\bibnamefont {{The LIGO Scientific
  Collaboration}}}, \bibinfo {author} {\bibnamefont {{the Virgo
  Collaboration}}},\ and\ \bibinfo {author} {\bibnamefont {{the KAGRA
  Collaboration}}},\ }\href {https://doi.org/10.5281/zenodo.5636816} {\bibinfo
  {title} {Gwtc-3: Compact binary coalescences observed by ligo and virgo
  during the second part of the third observing run — o1+o2+o3 search
  sensitivity estimates}} (\bibinfo {year} {2021}{\natexlab{e}})\BibitemShut
  {NoStop}%
\bibitem [{\citenamefont {{The LIGO Scientific Collaboration}}\ \emph
  {et~al.}(2021{\natexlab{f}})\citenamefont {{The LIGO Scientific
  Collaboration}}, \citenamefont {{the Virgo Collaboration}},\ and\
  \citenamefont {{the KAGRA Collaboration}}}]{o3lvkinjections}%
  \BibitemOpen
  \bibfield  {author} {\bibinfo {author} {\bibnamefont {{The LIGO Scientific
  Collaboration}}}, \bibinfo {author} {\bibnamefont {{the Virgo
  Collaboration}}},\ and\ \bibinfo {author} {\bibnamefont {{the KAGRA
  Collaboration}}},\ }\href {https://doi.org/10.5281/zenodo.5546676} {\bibinfo
  {title} {Gwtc-3: Compact binary coalescences observed by ligo and virgo
  during the second part of the third observing run — o3 search sensitivity
  estimates}} (\bibinfo {year} {2021}{\natexlab{f}})\BibitemShut {NoStop}%
\end{thebibliography}%
\end{document}